\documentclass[10pt,conference]{IEEEtran}

\IEEEoverridecommandlockouts

\usepackage{cite}
\usepackage{amsmath,amssymb,amsfonts}
\usepackage{algorithmic}
\usepackage{graphicx}
\usepackage[dvipsnames]{xcolor}
\usepackage[final]{microtype}
\usepackage[italic]{mathastext}
\usepackage{libertine}
\usepackage[T1]{fontenc}
\usepackage{textcomp}
\usepackage[varqu,varl]{zi4}
\usepackage[all]{nowidow}
\usepackage[keeplastbox]{flushend}
\usepackage{fancyhdr}
\usepackage{enumitem}
\usepackage[colorlinks=true, linkcolor=blue, citecolor=blue, urlcolor=blue]{hyperref}
\usepackage{subcaption}
\usepackage{booktabs}
\usepackage{multirow}
\usepackage{multicol}
\usepackage[table]{xcolor}
\usepackage{pifont}
\newcommand{\cmark}{\ding{51}}%
\newcommand{\xmark}{\ding{55}}%
\usepackage{fancybox}  
\usepackage[breakable]{tcolorbox}
\usepackage{tcolorbox}  

\def\BibTeX{{\rm B\kern-.05em{\sc i\kern-.025em b}\kern-.08em
    T\kern-.1667em\lower.7ex\hbox{E}\kern-.125emX}}

\fancypagestyle{firstpage}{
    \fancyhf{}
    
    \fancyhead[C]{\vspace{10pt}\normalsize{To appear at 2025 IEEE International Symposium on Workload Characterization \\\vspace{-25pt}} 
    \fancyfoot[C]{\thepage}
    }
}

\begin{document}

\title{ZKProphet: Understanding Performance of Zero-Knowledge
Proofs on GPUs
}

\author{
    \IEEEauthorblockN{Tarunesh Verma, Yichao Yuan, Nishil Talati, Todd Austin}
    \IEEEauthorblockA{
        \textit{Computer Science and Engineering, University of Michigan, USA} \\
        \{tarunesh, yichaoy, talatin, austin\}@umich.edu
        }
}

\maketitle

\thispagestyle{firstpage}
\pagestyle{plain}

\begin{abstract}\label{section:abstract}
Zero-Knowledge Proofs (ZKP) are protocols which construct cryptographic proofs to demonstrate knowledge of a secret input in a computation without revealing any information about the secret.
ZKPs enable novel applications in private and verifiable computing such as anonymized cryptocurrencies and blockchain scaling and have seen adoption in several real-world systems. 
Prior work has accelerated ZKPs on GPUs by leveraging the inherent parallelism in core computation kernels like Multi-Scalar Multiplication (MSM).
However, we find that a systematic characterization of execution bottlenecks in ZKPs, as well as their scalability on modern GPU architectures, is missing in the literature. 

This paper presents ZKProphet, a comprehensive performance study of Zero-Knowledge Proofs on GPUs. 
Following massive speedups of MSM, we find that ZKPs are bottlenecked by kernels like Number-Theoretic Transform (NTT), as they account for up to 90\% of the proof generation latency on GPUs when paired with optimized MSM implementations.
Available NTT implementations under-utilize GPU compute resources and often do not employ architectural features like asynchronous compute and memory operations. 
We observe that the arithmetic operations underlying ZKPs execute exclusively on the GPU's 32-bit integer pipeline and exhibit limited instruction-level parallelism due to data dependencies.
Their performance is thus limited by the available integer compute units.
While one way to scale the performance of ZKPs is adding more compute units, we discuss how runtime parameter tuning for optimizations like precomputed inputs and alternative data representations can extract additional speedup. 
With this work, we provide the ZKP community a roadmap to scale performance on GPUs and construct definitive GPU-accelerated ZKPs for their application requirements and available hardware resources.
\end{abstract}

\section{Introduction} \label{section:introduction}

\emph{Zero-Knowledge Proofs} are cryptographic protocols in which one party (the $Prover$) produces a proof of knowledge $\pi$ to convince another party (the $Verifier$) that the $Prover$ has correctly performed the computation $f(x, w) = y$, where $f$ is a public function, $x$ is a public input, and $w$ is a private input (aka "witness") known only to the $Prover$. 
The proof $\pi$ does not reveal any information about $w$ other than the $Prover$'s knowledge of $w$.
ZKPs have been adopted in real-world systems for private cryptocurrencies, computation outsourcing, and blockchain rollups \cite{app-filecoin, app-zcash, app-celo, app-loopring, app-ing}, and have been studied for privacy in network middleboxes \cite{app-zkmb, app-zombie}, verifiable machine learning \cite{app-verif-ml}, verifiable homomorphic encryption \cite{app-greco}, and improved server authentication \cite{app-nope}. 
Proof generation latency scales with the complexity of the computation $f$ and takes several minutes on modern CPUs, while verification is constant-time and requires a few milliseconds \cite{zkp-groth}. 
Accelerating $Prover$ is thus paramount to wider adoption of ZKPs \cite{app-zkmb, app-zombie, app-zkdl, app-nope}.

\begin{figure}[t]
    \centering
	\includegraphics[width=0.45\textwidth, trim={0cm 0cm 0cm 0cm}]{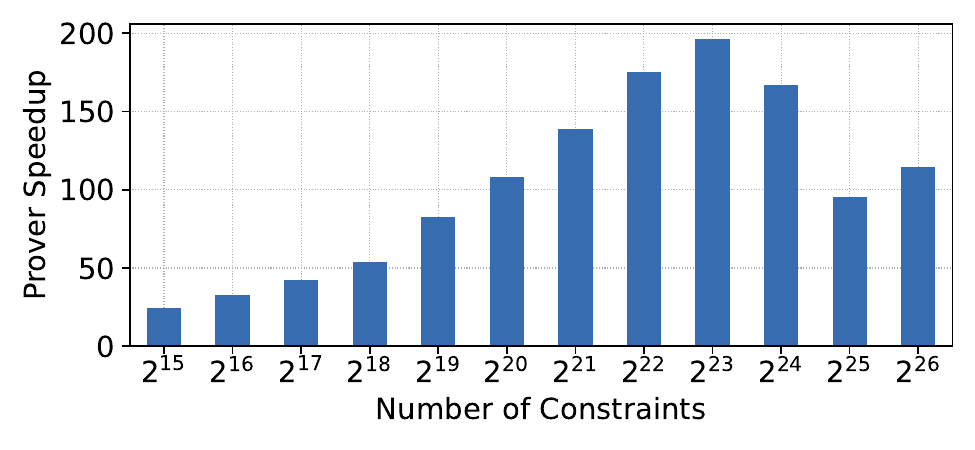}
	\vspace{-0.3cm}
	\caption
	{
	    {Speedup of ZKP GPU implementations over CPU.}
	}
	\label{figure:proof-gpu-speedup}
    \vspace{-0.65cm}
\end{figure}

Given the data-parallel nature of proof generation, GPUs have emerged as attractive platforms for accelerating ZKPs \cite{gpu-bellperson, gpu-cuzk, gpu-distmsm, gpu-elasticmsm, gpu-gzkp, gpu-icicle, gpu-sppark, gpu-ymc, opt-zprize}. 
The underlying Multi-Scalar Multiplication (MSM) and Number-Theoretic Transform (NTT) kernels, accounting for $>$95\% of the workload, are highly parallelizable compute-intensive tasks which can benefit from GPU implementations.
This is evident in Figure \ref{figure:proof-gpu-speedup}, which shows that GPU-accelerated ZKPs are up to $\sim$200x faster than CPU baselines. 
The number of constraints refers to the number of inputs to the kernels and is determined by the complexity of the computation $f$ being proved. 

While prior work has achieved significant speedups for individual kernels on GPUs, we observe that a systematic characterization of end-to-end proof generation on GPUs, which studies the performance bottlenecks and scalability on modern GPU architectures, is missing.
Moreover, ZKP frameworks for end-users \cite{cpu-arkworks, gpu-icicle, gpu-bellperson, cpu-circom, cpu-xjsnark} offer their own implementations of the underlying computation kernels with varying levels of performance and abstract away implementation details behind high-level interfaces.
This furthers the gap between end-users achieving the best performance for their ZKP workloads.

\begin{figure}[t]
    \centering
	\includegraphics[width=0.48\textwidth, trim={0cm 0cm 0cm 0cm}]{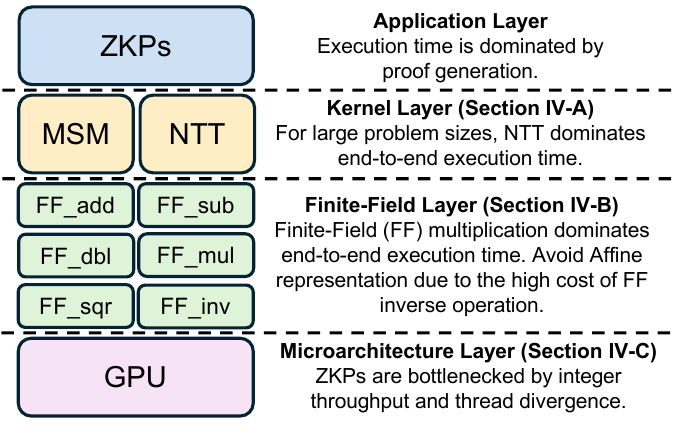}
	\caption
	{
	    Overview of our performance analysis.
	}
    \vspace{-0.3cm}
	\label{figure:zkprophet-analysis}
\end{figure}

To address these limitations, we propose ZKProphet, a comprehensive performance study of GPU-accelerated ZKPs.
Figure \ref{figure:zkprophet-analysis} shows an overview of our analysis. 
We focus on publicly available implementations of core computation kernels compatible with the Groth16 ZKP \cite{zkp-groth}, chosen for its succinct proofs and sub-millisecond verification time. % regardless of the proof generation latency. 
Since proof generation is the computationally intensive task accelerated on GPUs (Figure \ref{figure:zkprophet-analysis}) while verification is constant time, this paper analyzes the $Prover$ and uses this term interchangeably with ZKP.

We evaluate state-of-the-art ZKP libraries on several generations of GPUs and find that the performance of MSM has far outpaced that of NTT (Figure~\ref{figure:zkprophet-analysis}).
MSM, traditionally $\sim$70\% of the runtime \cite{gpu-cuzk, opt-zprize}, has been the primary focus of prior acceleration efforts.
We find that in optimized $Prover$s, NTT contributes up to 90\% of the ZKP latency.

Our study finds that NTT implementations often under-utilize the available GPU resources and do not leverage modern architectural capabilities to hide latencies, while optimized MSM implementations are tailored to specific GPUs and offer sub-optimal performance on different targets. 
Furthermore, the number of constraints in the computation determines the choice of the ideal MSM and NTT implementations for end-to-end workloads.
%These disparate MSM and NTT libraries offer limited interoperability with each other and with end-to-end ZKP frameworks like \emph{arkworks} \cite{cpu-arkworks}. 
These disparate libraries offer limited interoperability with each other and with end-to-end ZKP frameworks like \emph{arkworks} \cite{cpu-arkworks}. 
We therefore find opportunities to unify different ZKP frameworks to enable plug-and-play solutions for developers who can leverage the best tools for their ZKP applications without needing to understand underlying cryptographic primitives and GPU programming.

We subsequently take a quantitative approach to characterize the performance of the underlying integer arithmetic operations in MSM and NTT kernels.
These operations are performed in a finite field i.e., the results are bound by a large prime number. 
We find that finite-field multiplication is the primary component in MSM and NTT kernels (Figure~\ref{figure:zkprophet-analysis}).
This operation exhibits limited Instruction-Level Parallelism and its performance is limited by the 32-bit integer execution units available on the GPU. 
Traditional instruction latency-hiding techniques, such as increasing the number of threads, often degrade ZKP performance.

Analyzing several generations of modern NVIDIA GPUs, we observe that ZKP performance improves primarily by adding Streaming Multiprocessors (SMs), as the 32-bit integer performance per SM has remained constant (Figure~\ref{figure:zkprophet-analysis}).
Moreover, we observe that architectural improvements in newer GPUs, specifically greater memory bandwidth and capacity, increased shared-memory, and high-throughput execution units are not exploited by existing implementations. 
We show how intelligently tuning runtime parameters can improve performance on newer GPUs. 

\section{Background on ZKP} \label{section:background}
In a ZKP application, the $Prover$ produces a proof and transmits it to the $Verifier$.
Compact proofs with efficient verifications have low network and storage requirements and can enable ZKP technology at scale \cite{zkp-recursive-circom, app-hermez, app-celo}.
In this paper, we focus on the Groth16 \cite{zkp-groth} ZKP, as these proofs are less than 200 bytes and can be verified in less than 1 ms, making them orders of magnitude more efficient than other ZKPs \cite{asic-nocap, app-zombie, zkp-orion, zkp-stark}. 
Groth16 is supported by state-of-the-art ZKP libraries \cite{cpu-arkworks, cpu-libsnark, cpu-jsnark, cpu-xjsnark, cpu-circom, cpu-dizk, cpu-gnark, gpu-bellperson, gpu-cuzk} and has seen adoption in several real-world applications \cite{app-zcash, app-darkforest, app-loopring, app-filecoin, app-hermez, app-celo, app-semaphore}.
In Groth16, MSM and NTT kernels dominate the end-to-end execution time by more than 90\%~\cite{gpu-cuzk}.
Additionally, MSM and NTT are used in several other ZKP protocols like Marlin \cite{zkp-marlin} , PLONK \cite{zkp-plonk} (and its variants), Sonic \cite{zkp-sonic}, Bulletproofs \cite{zkp-bulletproofs} , HALO \cite{zkp-recursive-halo-inf}, Orion \cite{zkp-orion}, Virgo \cite{zkp-virgo}, STARK \cite{zkp-stark}, Aurora \cite{zkp-aurora}, and Ligero \cite{zkp-ligero}.

\begin{figure}[t]
    \centering
	\includegraphics[width=0.50\textwidth, trim={1.5cm 10.5cm 15.25cm 3.5cm}, clip]{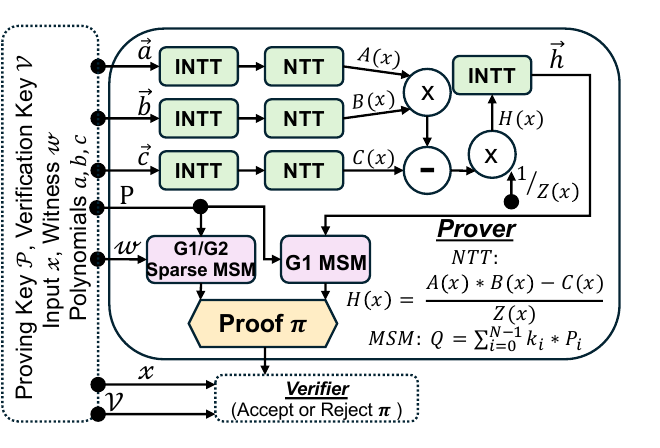}
	\vspace{-0.7cm}
	\caption
	{
	    {Groth16 Protocol showing $Prover$ computations.}
	}
	\label{figure:proof-protocol}
    \vspace{-0.3cm}
\end{figure}

\begin{figure*}[t]
    \centering
	\includegraphics[width=0.90\textwidth, trim={1.5cm 8.5cm 1.5cm 3.7cm}, clip]{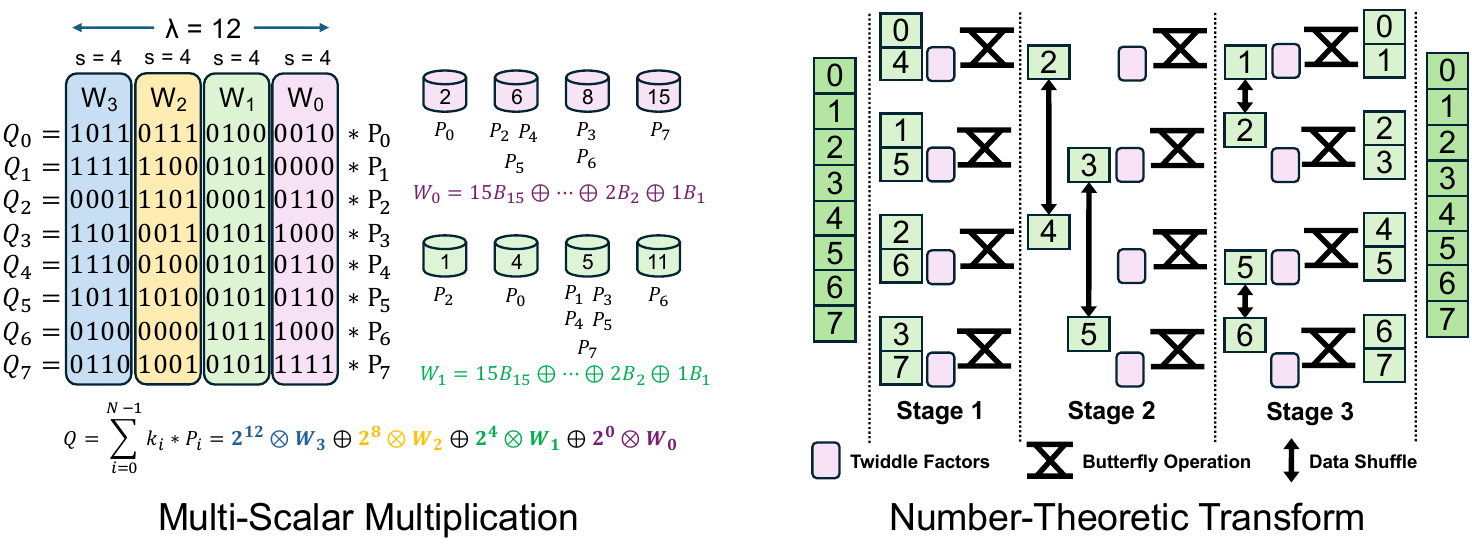}
	\vspace{-0.2cm}
	\caption
	{
	    {(a) Pippenger's Bucket Algorithm for MSM and (b) Cooley-Tukey Algorithm for NTT.}
	}
	\label{figure:msm-ntt}
    \vspace{-0.5cm}
\end{figure*}

Figure \ref{figure:proof-protocol} shows an overview of the Groth16 ZKP.
The application and its public and private inputs, $x$ and $w$, are encoded into a set of polynomials $\vec{a}$, $\vec{b}$, $\vec{c}$, and $\vec{Z}$, upon which a series of NTT operations is performed.
These polynomials consist of large integers (e.g., 256-bit). 
The resultant polynomial and the private input $w$ are combined with a Proving Key $\mathcal{P}$, which also consists of large integers (e.g., 377-bit), using MSM operations to generate the proof $\pi$.
The number of elements in the polynomials and the proving key, referred to as the number of constraints, is determined by the complexity of the application.
We refer the reader to~\cite{zkp-groth} for additional details on the proof system.

The integers in MSM and NTT are elements in a \emph{finite field}.
Briefly, a finite field $\mathbb{F}_p$ is a set of integers between 0 and a large prime number $p$ (i.e., the field modulus) which supports arithmetic operations like addition, subtraction, multiplication, and inverse. 
These operations are performed modulo $p$ i.e., the results of field operations always lie between $[0, p)$. 
For NTT, the inputs are integers in a finite field $\mathbb{F}_r$.
For MSM, the elements of the Proving Key $\mathcal{P}$ are points on an elliptic curve chosen for cryptographic security and performance properties.
These points on the elliptic curve consists of 2--4 coordinates, where each coordinate is a large integer in a finite field $\mathbb{F}_q$. 
The implementations studied in this work support BLS12-377 and BLS12-381 elliptic curves and associated finite fields. 

Since the large integers are longer than the word size of modern GPUs, they are represented using word-sized \emph{limbs}: a 377-bit integer can be represented using 12 32-bit limbs and the field operations are performed on these limbs. 

%%%%%%%%%%%%%%%%%%%%%%%%%%%%%%%%%%%%%%%%%%%%%%%%%%%%%%%%%%%%%%%%%%%%%%%%%%%%%%%%%%%%%%%%%%%%%%%%%%%%%%%%%%%%%%%%%%%%%

\subsection{Multi-Scalar Multiplication (MSM)} \label{subsection:background-msm}
MSM sums up the dot product between elliptic curve points and scalar integers: $ Q = \sum_{i = 0}^{N - 1} k_i \cdot P_i $.
The \emph{scale}, $N$, is determined by the complexity of the computation for which a proof is being generated.
For real-world applications, $N$ is on the order of millions.
In Groth16, MSM calculates the polynomial commitments to ensure an honest $Prover$ and enable succinct verification \cite{zkp-groth}. 

To multiply a point $P_i$ with a scalar $k_i$, $P_i$ is added to itself $k_i$ times using Point-Addition ($PADD$) and Point-Doubling ($PDBL$) formulae.
$PADD$ and $PDBL$ are composed of a series of modular arithmetic operations on the underlying integer coordinates as determined by different forms of elliptic curve points. 
Common forms include Affine with 2 coordinates $(x, y)$, Jacobian with 3 coordinates $(X, Y, Z)$, and XYZZ with 4 coordinates $(X, Y, ZZ, ZZZ)$.  
\cite{ec-hyperelliptic-db} provides a list of $PADD$ and $PDBL$ algorithms for various representations, and we explore these forms in \S\ref{results-ff-level}. 

MSM is performed using Pippenger's Algorithm \cite{opt-pippenger}, shown in Figure \ref{figure:msm-ntt}(a).
A $\lambda$-bit scalar is split into $w$ windows of $s$-bits each. 
Within each window, the $s$-bit scalar has $2^s$ values (organized as buckets).
The elliptic curve points within the window are placed into buckets with matching scalar values, and the buckets are then summed up ($PADD$) in the \emph{Bucket Accumulation} process. 
Each bucket sum is then multiplied by the corresponding bucket value and all the bucket values are added up in the \emph{Bucket Reduction} process.
This weighted sum is calculated with the Sum-of-Sums algorithm \cite{gpu-ymc} for each window, leaving us with $w$ partial sums. 
Finally, the weighted sum is performed on the window sums in the \emph{Window Reduction} process with $PADD$ and $PDBL$ operations to get the final result. 

Pippenger's Algorithm is highly parallel in nature. 
\emph{Bucket Accumulation} and \emph{Bucket Reduction} can be performed for each window independently, with one thread typically assigned to one bucket. 
The $N$ points and scalars processed within each window can be split into multiple sub-tasks, where $n < N$ points per window can be processed in parallel and then combined. 
\emph{Window Reduction} is serial and often performed on the CPU \cite{gpu-ymc, gpu-bellperson, gpu-gzkp, asic-pipezk}. 
Numerous prior works have accelerated MSM on GPUs \cite{gpu-bellperson, gpu-cuzk, gpu-distmsm, gpu-elasticmsm, gpu-gzkp, gpu-icicle, gpu-sppark, gpu-ymc} to achieve 2-3 orders of magnitude speedup over CPU implementations for the dominant G1 MSM.
G2 MSM is performed in parallel on CPU \cite{asic-pipezk}.

%%%%%%%%%%%%%%%%%%%%%%%%%%%%%%%%%%%%%%%%%%%%%%%%%%%%%%%%%%%%%%%%%%%%%%%%%%%%%%%%%%%%%%%%%%%%%%%%%%%%%%%%%%%%%%%%%%%%%
\subsection{Number-Theoretic Transform (NTT)} \label{subsection:background-ntt}

NTT computation is the Fast Fourier Transform for elements in a finite field.
NTT maps a vector of field elements $a = [a_0, a_1, ... a_n]$ to $A = [A_0, A_1, ... A_n]$, where $A_i = \sum_{j=0}^{n-1} a_i \omega^{ij}$. 
$\omega$ is the primitive $n$-th root of unity in the finite field, and the different powers of $\omega$ are known as the twiddle factors. 

Operations such as multiplication and addition on coefficients of polynomials are convolution operations with $O(n^2)$ complexity.
NTT transforms the polynomial coefficients to enable element-wise operations, with an overall complexity of $O(nlogn)$. 
Two polynomials can be multiplied by first transforming the coefficients using NTTs, then performing an element-wise multiplication, followed by inverse NTTs.
As shown in Figure \ref{figure:proof-protocol}, the Groth16 $Prover$ performs a series of forward and inverse NTTs interspersed with element-wise operations to calculate the polynomial $\vec{h}$.

Figure \ref{figure:msm-ntt}(b) shows the radix-2 Cooley-Tukey algorithm \cite{ntt-cooley-tukey}. 
The butterfly operation on elements 0 and 4 calculates $A_0 = A_0 + A_4\cdot\omega^{0}$ and $A_0 = A_0 - A_4\cdot\omega^{0}$ using modular addition, subtraction, and multiplication.
A scale $N$ NTT contains $N/2$ butterfly operations and $log_2(N)$ stages.  
The input elements are shuffled after each stage. 

The NTT algorithm is also highly parallelizable. 
In a typical GPU implementation \cite{gpu-bellperson, gpu-cuzk}, each thread performs the Butterfly Operation on a pair of elements.
The input vector is divided among several blocks, and each block with r threads performs a radix-2r Cooley-Tukey algorithm. 
The sizes of the blocks are determined by the capacity of the shared memory, which is used for data shuffles and storing the precomputed twiddle factors.
Different stages can be processed in batches.
Prior GPU acceleration efforts for ZKPs \cite{gpu-bellperson, gpu-sppark, gpu-cuzk, gpu-gzkp} have achieved 1-2 orders of magnitude of speedup over CPU implementations.
In our analysis, INTT and NTT exhibit a similar performance profile, and in the rest of this paper we represent them as NTT. 

\section{Experimental Methodology}\label{section:methodology}

\subsection{ZKP Libraries}\label{subsection:methodology-libraries}
\begin{table}[htbp]
  \centering
  \scriptsize
  \captionsetup{justification=centering}
  \begin{tabular}{|c|c|c|c|c|}
    \hline
    \textbf{Library} & \textbf{Platform} & \textbf{MSM} & \textbf{NTT} & \textbf{Groth16 $Prover$}\\
    \hline
    arkworks\cite{cpu-arkworks}     & CPU & \cmark & \cmark & \cmark \\
    \hline
    bellperson\cite{gpu-bellperson} & GPU & \cmark & \cmark & \cmark \\
    \hline 
    sppark\cite{gpu-sppark}         & GPU & \cmark & \cmark & \xmark \\
    \hline 
    cuzk\cite{gpu-cuzk}             & GPU & \cmark & \cmark & \cmark \\
    \hline 
    yrrid\cite{gpu-ymc}             & GPU & \cmark & \xmark & \xmark \\
    \hline 
    ymc\cite{gpu-ymc}               & GPU & \cmark & \xmark & \xmark \\
    \hline 
  \end{tabular}
  \caption{ZKP libraries evaluated in this work.}
  \label{table:zkp-libraries}
  \vspace{-0.3cm}
\end{table}

Table \ref{table:zkp-libraries} lists the ZKP libraries evaluated in this work.
\textbf{\emph{arkworks}} \cite{cpu-arkworks} is a CPU-based Rust framework for developing ZKPs and supports a variety of proof systems (including Groth16) and supports for a variety of elliptic curves, finite fields, and point representations. 

\textbf{\emph{bellperson}} \cite{gpu-bellperson} is a GPU library for Groth16, with OpenCL/CUDA implementations of Pippenger's MSM Algorithm and radix-2r Cooley-Tukey NTT. 
\textbf{\emph{sppark}} \cite{gpu-sppark} is a GPU library with optimized implementations of MSM and NTT kernels which uses \emph{arkworks} to instantiate different finite fields. 
\textbf{\emph{cuZK}} \cite{gpu-cuzk} accelerates the Groth16 protocol with its own framework using novel parallelization techniques to achieve significant speedup over baseline CPU implementations.
\textbf{\emph{yrrid}} \cite{gpu-yrrid} is a GPU library from the ZPrize competition \cite{opt-zprize}, an industry-sponsored effort to accelerate a batch of MSMs with scale $2^{26}$ on GPUs for the BLS12-377 elliptic curve.
\textbf{\emph{ymc}} \cite{gpu-ymc} augments \emph{yrrid} with optimized finite-field routines and workload decomposition techniques for additional performance gains.

We restrict our analysis to the above mentioned libraries because of their high performance, functionality, and compatibility with ZKP computations. 
\textbf{\emph{ICICLE}} \cite{gpu-icicle} is another GPU acceleration library for ZKPs, but the performance analysis opportunities are limited as the GPU implementations aren't open-source.
\textbf{\emph{GPU-NTT}} by Ozcan et al. \cite{ntt-ozcan} provides optimized NTT algorithms with performance improvements over \emph{sppark}. 
However, the publicly available implementation isn't yet compatible with the finite fields required for ZKPs. 

\subsection{Software and Hardware Infrastructure}\label{subsection:methodology-hw-sw}
We utilize \emph{arkworks} to generate test cases and instantiate the underlying cryptographic primitives like elliptic curves and finite fields and measure CPU baselines.
\emph{arkworks} is an open-source framework supporting a variety of proof systems, cryptographic primitives, and computation kernels and is adopted in industry-sponsored ZKP acceleration efforts like \cite{opt-zprize}.
The GPU kernels and additional microkernels for performance characterization are written in C++ and CUDA and compiled using CUDA Toolkit 12.8 on Ubuntu 22.04. 
\emph{cuZK} was tested using CUDA Toolkit 11.5 as that is the latest version supported by the implementation.
We use NVIDIA Nsight Compute 2025.1 to profile the applications and analyze the performance on GPUs and measure the execution latency using cycle counters.
We measure the energy consumption of CPU and GPU implementations using Zeus \cite{zeus-energy}. 

The CPU baselines are evaluated on a dual socket server with AMD EPYC 7742 64-Core Processors and 2 TB RAM. 
GPU studies are primarily conducted on NVIDIA A40 GPU with 48GB of memory.
In \S\ref{results-diff-gpus}, we perform additional analysis of key finite-field operations on several generations of NVIDIA GPUs: Volta V100, Turing T4, Ampere RTX 3090 and A100, Ada Lovelace L4 and L40S, and Hopper H100.

\subsection{Key Research Questions}
As described in \S\ref{section:background}, GPUs are suitable targets for accelerating the MSM and NTT algorithms. 
Several optimized GPU implementations for MSM and NTT have been proposed recently in industry and academia \cite{gpu-bellperson, gpu-cuzk, gpu-gzkp, gpu-elasticmsm, gpu-distmsm, gpu-icicle, gpu-sppark, gpu-ymc}.
While these implementations build upon Pippenger's and radix-2r Cooley-Tukey techniques, they differ in their choices of algorithmic optimizations, elliptic curve point representations, parallelization techniques, and other GPU implementation details, introducing varying computation and storage overheads. 

Therefore, we first seek to understand \textbf{which implementations are fastest at different input sizes and why}.
This insight is crucial for application developers who seek to maximize ZKP performance without diving into underlying cryptography and GPU architectural details. 
We then evaluate \textbf{the overall $Prover$ latency using the fastest kernels} (\S\ref{results-kernel-level}) to find where the bottlenecks lie.
We study the \textbf{performance and microarchitectural execution of the modular finite-field operations in MSM and NTT} (\S\ref{results-ff-level} and \S\ref{results-sass-level}) to discover key optimization targets.
Finally, we ask \textbf{how the performance of finite-field operations has evolved over subsequent generations of NVIDIA GPU architectures} (\S\ref{results-diff-gpus}) to understand the effect of GPU architecture scaling on the ZKP workload.
\section{ZKProphet: A Performance Deep-Dive} \label{section:results}

%%%%%%%%%%%%%%%%%%%%%%%%%%%%%%%%%%%%%%%%%%%%%%%%%%%%%%%%%%%%%%%%%%%%%%%%%%%%%%%%
%%%%%%%%%%%%%%%%%%%%%%%%%%%%%%%%%%%%%%%%%%%%%%%%%%%%%%%%%%%%%%%%%%%%%%%%%%%%%%%%
\subsection{Analysis at the Kernel Layer} \label{results-kernel-level}

\definecolor{colorSppark}{HTML}{386cb0}
\definecolor{colorYrrid}{HTML}{FFD23F}
\definecolor{colorYmc}{HTML}{DB5461}
\definecolor{colorBellperson}{HTML}{7fc97f}
\definecolor{colorCuzk}{HTML}{fdc086}

\begingroup
    \renewcommand{\arraystretch}{1.4}
    \begin{table}[htbp]
      \centering
      \scriptsize
      \begin{tabular}{|c|c|c|c|c|}
        \hline
        %------ header row 1 ------
        % multirow for Col 1 header + first data‐cell
        \multirow{2}{*}{\textbf{Scale}}
          & \multicolumn{2}{c|}{\textbf{MSM}}
          & \multicolumn{2}{c|}{\textbf{NTT}} \\
        \cline{2-5}
        %------ header row 2 ------
        % (the blank in column 1 is “eaten” by the \multirow)
          & \textbf{\shortstack{Speedup ($\times$)\\over CPU}} & \textbf{\shortstack{Fastest\\Library}} & \textbf{\shortstack{Speedup ($\times$)\\over CPU}} & \textbf{\shortstack{Fastest\\Library}} \\
        \hline
        %------ first data row (Cell11 is merged up into the header) ------
        $2^{15}$ & 34.1  & \cellcolor{colorSppark!70}sppark  &  12.5 & \cellcolor{colorBellperson!70}bellperson \\
        \hline
        $2^{16}$ & 52.5  & \cellcolor{colorSppark!70}sppark  &  12.3 & \cellcolor{colorBellperson!70}bellperson \\
        \hline
        $2^{17}$ & 69.7  & \cellcolor{colorSppark!70}sppark  &  14.8 & \cellcolor{colorBellperson!70}bellperson \\
        \hline
        $2^{18}$ & 78.1  & \cellcolor{colorSppark!70}sppark  & 20.4 & \cellcolor{colorCuzk!70}cuzk    \\
        \hline
        $2^{19}$ & 127.5 & \cellcolor{colorSppark!70}sppark  & 27.9 & \cellcolor{colorCuzk!70}cuzk    \\
        \hline
        $2^{20}$ & 176.1 & \cellcolor{colorSppark!70}sppark  & 35.4 & \cellcolor{colorCuzk!70}cuzk    \\
        \hline
        $2^{21}$ & 254.1 & \cellcolor{colorYrrid!70}yrrid & 45.0 & \cellcolor{colorCuzk!70}cuzk    \\
        \hline
        $2^{22}$ & 408.1 & \cellcolor{colorYmc!70}ymc    & 50.6 & \cellcolor{colorCuzk!70}cuzk    \\
        \hline
        $2^{23}$ & 589.4 & \cellcolor{colorYmc!70}ymc    & 50.3 & \cellcolor{colorCuzk!70}cuzk    \\
        \hline
        $2^{24}$ & 693.2 & \cellcolor{colorYmc!70}ymc    & 40.5 & \cellcolor{colorBellperson!70}bellperson \\
        \hline
        $2^{25}$ & 754.3 & \cellcolor{colorYmc!70}ymc    & 20.4 & \cellcolor{colorBellperson!70}bellperson \\
        \hline
        $2^{26}$ & 799.5 & \cellcolor{colorYmc!70}ymc    & 24.3 & \cellcolor{colorBellperson!70}bellperson \\
        \hline
      \end{tabular}
      \caption{Speedup over CPU for the fastest MSM and NTT implementations at different input sizes (Scale).}
      \label{table:msm-ntt-speedup}
      \vspace{-0.5cm}
    \end{table}
\endgroup

We evaluate the libraries listed in Table \ref{table:zkp-libraries} and report the speedups of the fastest MSM and NTT implementations over CPU baselines at different scales in Table \ref{table:msm-ntt-speedup}.
Scale refers to the input size for MSM and NTT and is determined by the application circuit for which a proof is generated.
The table shows that \textbf{\textit{there is no single implementation that performs best at different scales.}}

For MSM, \emph{sppark} \cite{gpu-sppark} is the fastest implementation at scales up to $2^{20}$. 
The implementation achieves acceleration through the XYZZ point representation, sorting the Pippenger buckets by the number of points for balanced workload distribution across threads, and minimizing the number of SASS instructions through tailored finite-field routines.
For larger scales, \emph{ymc} \cite{gpu-ymc} offers the highest speedup. 
In addition to the optimizations employed by \emph{sppark}, \emph{ymc} exploits (1) signed-digit endomorphism to halve the number of buckets, (2) pre-computed window weights to minimize Bucket Reductions, and (3) decomposition of large-scale MSMs into smaller MSMs to overlap compute with data transfer.
The implementation is tailored to problem scales for the Z-Prize competition \cite{opt-zprize}, and the pre-processing required for these optimizations is expensive at smaller scales, taking up to 30\% of the MSM compute time. 
\emph{ymc} is thus better for larger scales or large batches of MSMs.

For NTT, \emph{bellperson} \cite{gpu-bellperson} offers the highest speedup at scales $2^{15}$ to $2^{17}$. 
It implements the radix-256 Cooley-Tukey algorithm to combine up to 8 NTT stages into a single kernel launch. 
At larger scales, \emph{cuzk} \cite{gpu-cuzk} emerges as the fastest implementation, improving performance by reducing CPU--GPU data movement, storing precomputed twiddle factors in device memory, and coalescing GPU memory operations. 
For scales beyond $2^{23}$, \emph{cuZK} NTT reports Memory Allocation and Segmentation Fault errors.

\emph{bellperson} offers the best performance at scales beyond $2^{23}$.
However, the GPU workload distribution is not optimal.
A $2^{26}$ NTT uses 4 kernels - 3 radix-256 NTTs and 1 radix-2 NTT. 
The final radix-2 kernel has an imbalanced launch configuration of 16 million blocks of 2 thread each because of how the implementation is structured. 
This leads to critical underutilization of the GPU resources, specifically the Streaming Multiprocessor (SM) and its Sub-Partitions (SMSP), which run 32-thread warps in lockstep. 
Moreover, the kernel uses $<$5\% of the available device memory.
The reduced speedup at larger scales is shown in Figure \ref{figure:proof-gpu-speedup}.

\begin{figure}[!htb]
    \centering
    \includegraphics[width=0.5\textwidth, trim={0cm 0cm 0cm 0.5cm}, clip]{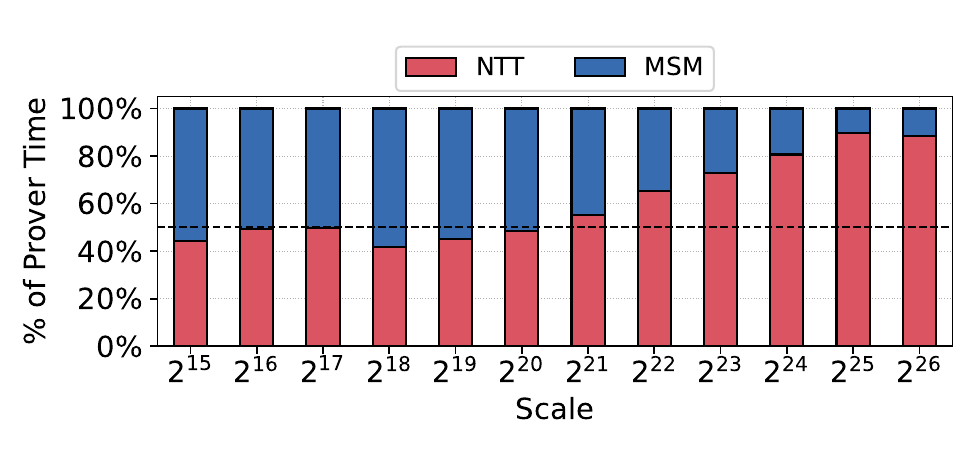}
	\vspace{-0.7cm}
    \caption{ZKP execution time breakdown into MSM and NTT.}% at different scales.}
    \label{figure:ntt-msm-pct-timing}
    \vspace{-0.5cm}
\end{figure}

\begin{figure}[!htb]
    \centering
	\includegraphics[width=0.5\textwidth, trim={0.3cm 0cm 0cm 0.5cm}, clip]{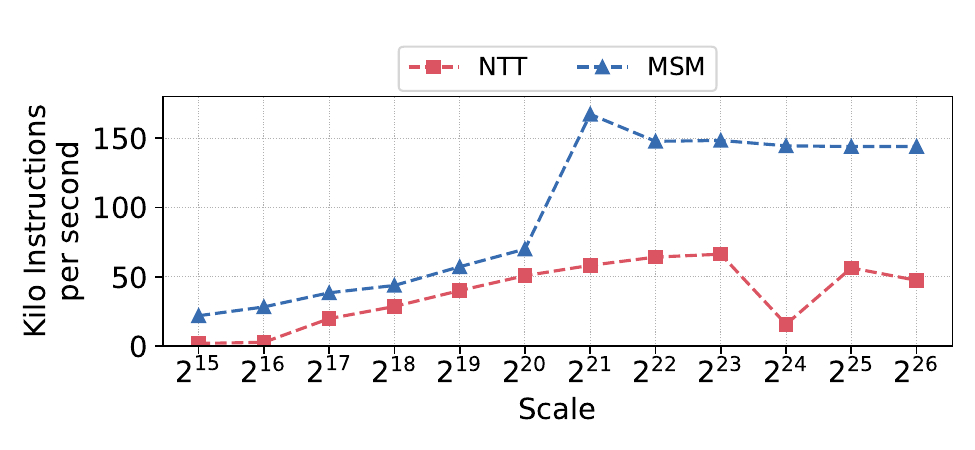}
	\vspace{-0.7cm}
	\caption
	{
	    {Kilo instructions per second executed by optimal MSM and NTT implementations for different scales.}
	}
	\label{figure:kilo-instructions-per-second}
    \vspace{-0.5cm}
\end{figure}

\begin{figure}[!htb]
    \centering
	\includegraphics[width=0.5\textwidth, trim={0cm 0cm 0cm 1cm}, clip]{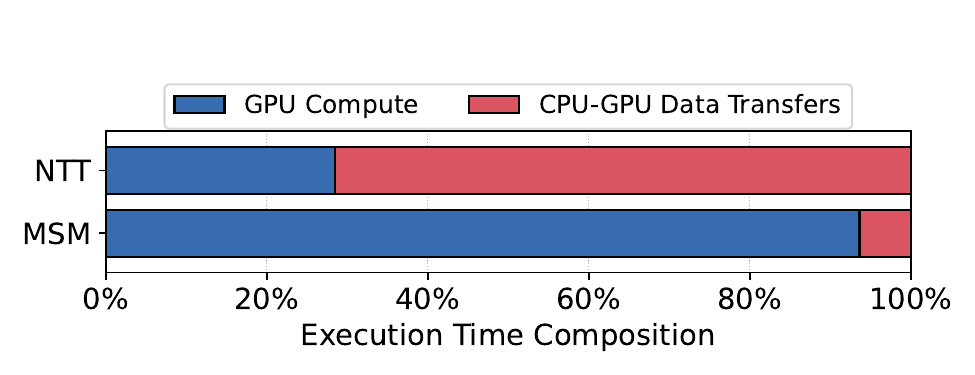}
	\vspace{-0.7cm}
	\caption
	{
	    {Percentage of execution time for on-device computation and CPU--GPU data transfer, averaged over scales $2^{23}$--$2^{26}$.}
	}
	\label{figure:ntt-msm-compute-vs-memory}
    \vspace{-0.5cm}
\end{figure}

Figure~\ref{figure:ntt-msm-pct-timing} shows the execution time breakdown of ZKP into MSM and NTT kernels at different scales.
This figure ignores other kernels as they are negligible with a time share of less than 5\% overall time.
The figure shows that even with modest proof sizes (up to $2^{20}$), NTT consumes $\sim$50\% of the proof generation time; this issue is exacerbated at larger proof sizes, where NTT contributes up to 91\% of $Prover$'s runtime. 
\textbf{\textit{Therefore, ZKP workloads are bottlenecked by NTT.}}

To further investigate, Figure~\ref{figure:kilo-instructions-per-second} compares the kilo instructions executed per second for the fastest MSM and NTT implementations across various problem scales. 
This metric reflects the GPU's instruction throughput, enabling a direct performance comparison between MSM and NTT.
As the problem scale increases, NTT executes significantly fewer instructions per unit time than MSM.
As noted in \S\ref{section:introduction}, GPU acceleration for ZKP workloads has largely focused on optimizing MSM kernels, driven in part by initiatives like the Z-Prize~\cite{opt-zprize}.
Our study highlights a \textbf{\textit{substantial opportunity to improve the performance of NTT kernels.}}

To understand this discrepancy, we compare the amount of time spent by both MSM and NTT on computing data on the GPU and CPU-GPU data transfers in Figure~\ref{figure:ntt-msm-compute-vs-memory}.
Optimized MSM implementations utilize asynchronous memory copies between CPU and the GPU and the GPU's global and shared memories to overlap data movement with compute. 
The latency of these memory operations are not hidden in NTT implementations like \emph{bellperson} \cite{gpu-bellperson}.
Furthermore, the $Prover$ performs seven NTT operations, each with multiple kernel launches.
We find that the on-device compute time of the butterfly operation is modest compared to the expensive CPU-GPU data transfers. 
This is shown in Figure~\ref{figure:ntt-msm-compute-vs-memory}, where the \textbf{\textit{fraction of the GPU compute time for NTT is much lower than CPU-GPU data transfer time than MSM.}}

% centered paragraph‐column of fixed width
\newcolumntype{P}[1]{>{\centering\arraybackslash}p{#1}}
\begin{table}[htbp]
  \centering
  % pick 40% of the text width for each GPU column:
  \newlength{\GpuColWidth}
  \setlength{\GpuColWidth}{0.2\linewidth}% 
  \begin{tabular}{|c|P{\GpuColWidth}|P{\GpuColWidth}|}
    \hline
    \multirow{2}{*}{\textbf{Scale}}
      & \multicolumn{2}{c|}{\textbf{CPU Energy Relative to GPU}} \\
    \cline{2-3}
      & \textbf{NTT} & \textbf{MSM} \\
    \hline
    16 &  2.74  &  2.74 \\ \hline
    18 &  3.08  &  9.06 \\ \hline
    20 &  3.21  & 27.59 \\ \hline
    22 &  3.31  &102.59 \\ \hline
    24 &  2.93  &236.90 \\ \hline
    26 &  3.62  &398.40 \\ \hline
  \end{tabular}
  \caption{CPU energy consumption normalized to GPU for NTT and MSM across different scales.}
  \label{table:energy_power_ratios}
\end{table}

We utilize Zeus \cite{zeus-energy} to evaluate the energy consumed by CPU and GPU implementations of NTT and MSM kernels at various scales. 
Table \ref{table:energy_power_ratios} reports the CPU energy consumption normalized to the GPU energy consumption for both kernels across varying scales. 
We observe that CPU-NTT consumes 3.1$\times$ more energy than GPU-NTT on average, while CPU-MSM can consume up to $\sim$400$\times$ more energy than GPU-MSM at scales of $2^{26}$. 
The energy efficiency of MSM on GPU stems primarily from the latency speedup of $\sim$800$\times$ compared to $\sim$50$\times$ for NTT, as discussed in Table \ref{table:msm-ntt-speedup}.
Additionally, the energy efficiency of GPU-MSM increases with the kernel scale due to well-optimized GPU implementations, whereas GPU-NTT executes short, bursty kernels often with sub-optimal launch parameters, as discussed earlier in this section.
These results underscore that \textbf{further NTT acceleration is crucial for improving energy efficiency} and facilitating ZKP deployment at scale. 

\begin{tcolorbox}[width=\linewidth]
\textbf{\emph{Key Takeaways:}}
  \begin{itemize}[leftmargin=*]
     \item No single MSM or NTT implementation is universally fastest across all problem scales.
     % \item The ideal set of algorithmic optimizations for MSM depends on the MSM scale.
     \item As MSM has been heavily optimized, NTT emerges as the dominant bottleneck.
     \item Current NTT implementations incur significant overhead from CPU-GPU data transfers.
  \end{itemize}
\end{tcolorbox}

%%%%%%%%%%%%%%%%%%%%%%%%%%%%%%%%%%%%%%%%%%%
%%%%%%%%%%%%%%%%%%%%%%%%%%%%%%%%%%%%%%%%%%%

\subsection{Analysis at the Finite-Field Layer}\label{results-ff-level}
The high-bitwidth integers used in MSM and NTT are elements in a finite field, and arithmetic operations are performed with modular reductions (\S\ref{section:background}).

\begin{figure}[t]
    \centering
	\includegraphics[width=0.5\textwidth, trim={1.4cm 0cm 1cm 0cm}, clip]{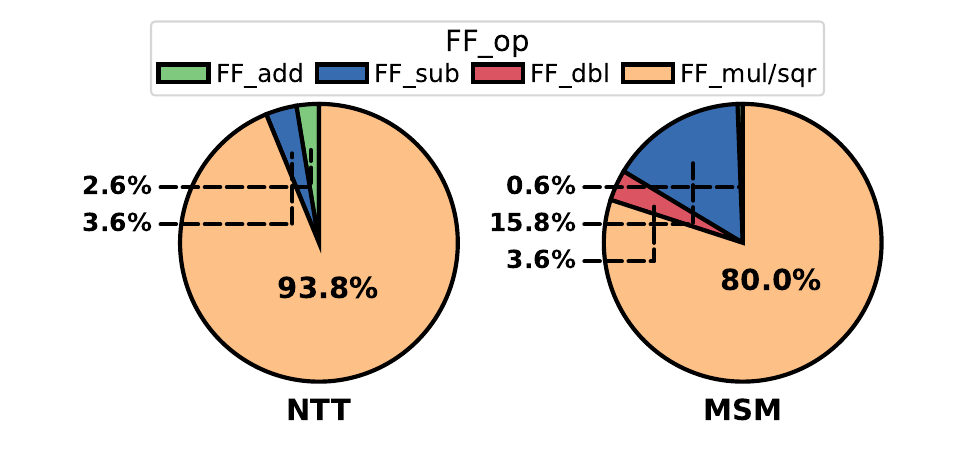}
	\vspace{-0.7cm}
	\caption
	{
	    {Breakdown of execution time into Finite-Field operations ($FF\_mul$ and $FF\_sqr$ have similar performance).}
	}
	\label{figure:ff-pct}
    %\vspace{-0.3cm}
\end{figure}

Figure \ref{figure:ff-pct} shows the percentages of the total execution times taken by the various field operations. 
NTT uses $FF\_add$, $FF\_sub$ and $FF\_mul$ in the butterfly operations and $FF\_mul$/$FF\_sqr$ to calculate the twiddle factors. 
MSM uses the field operations as part of the elliptic curve addition ($PADD$) and doubling ($PDBL$) functions. 
\textbf{\textit{$FF\_mul$ and $FF\_sqr$ exhibit similar performance profiles and are together responsible for 93.8\% and 80.0\% of the total execution time}} of NTT and MSM, respectively.
We characterize the performance of field operations on the CPU and the GPU using microbenchmarks designed to maximize GPU occupancy and limit expensive memory accesses. 

\begin{table}
  \centering
   \begin{tabular}{|c|c|c|c|c|c|}
    \hline
    \textbf{FF\_op} & \textbf{FF\_add} & \textbf{FF\_sub}
      & \textbf{FF\_dbl} & \textbf{FF\_mul}
      & \textbf{FF\_sqr} \\
    \hline
    \textbf{CPU} & 29 & 27 & 19 & 402 & 402 \\
    \hline
    \textbf{GPU} & 244 & 217 & 121 & 2656 & 2633 \\
    \hline
  \end{tabular}
  \caption{Execution latencies (in cycles) of finite-field operations on CPU and GPU.}
  \label{table:ff-ops-perf}
  \vspace{-0.5cm}
\end{table}

Table~\ref{table:ff-ops-perf} compares the \textit{latency of a single finite-field operation} on CPU and GPU.
CPUs can natively process 64-bit data elements compared to the 32-bit granularity of the GPU's integer units, and halving the number of limbs reduces the number of instructions.
After an $FF\_op$ completes, all limbs of the integer are sequentially compared against the field zero/modulus for underflow/overflow.
These conditional operations serialize the warp execution on the GPU (\textit{i.e.,} warp divergence), which further increases the latency gap with the CPU.
Despite a longer per-operation latency, GPUs can efficiently exploit the massive data-level parallelism in MSM and NTT to extract speedups shown in Figure~\ref{figure:proof-gpu-speedup}.
The NVIDIA A40 GPU features 84 streaming multiprocessors (SMs), each with 128 execution units capable of 32-bit integer operations, allowing it to run up to 10,752 threads in parallel.
In contrast, typical CPUs support only around 256 threads.
This stark difference underscores the importance of \textbf{\textit{leveraging data-level parallelism on massively parallel, high-throughput architectures to accelerate ZKP workloads effectively.}}

\subsubsection{Field Addition, Subtraction, and Doubling}\label{results-ff-add-sub-dbl}
Table \ref{table:ff-ops-perf} shows the GPU cycle latencies of $FF\_add$, $FF\_sub$ and $FF\_dbl$. 
This computation can be divided into two portions: a field operation (compute), control flow operations to determine if a reduction is necessary (branch), and a field operation to reduce the value (compute).
Our investigation shows that the branches determining conditional reduction make up 70.5\% of the overall execution latency.
In the absence of any branches, compute operations, $FF\_add$ and $FF\_sub$, require 72 cycles each.
The $FF\_add$ and $FF\_sub$ operations use the 32-bit \texttt{add\{c\}.cc} and \texttt{sub\{c\}.cc} PTX instructions, which are compiled into \texttt{IADD3} SASS instructions. 
$FF\_dbl$ efficiently doubles an element by left shifting each limb by $1$ and propagating carry-bits. 
It is implemented using bit-shifting and the \texttt{SHF} SASS instruction dominates $FF\_dbl$ operations. 
The latency of $FF\_dbl$ is reported in Table \ref{table:ff-ops-perf} and is lower than the latency of $FF\_add$. 

\subsubsection{Field Multiplication and Field Squaring} $FF\_mul$ multiplies two 32-bit field elements while $FF\_sqr$ multiplies an element with itself. 
As reported in Table \ref{table:ff-ops-perf}, $FF\_mul$ and $FF\_sqr$ require $\sim$10$\times$ more cycles to compute and conditionally reduce the result. 
These field operations are primarily composed of \texttt{IMAD} SASS instructions (70.8\% of the instruction mix). 
They are generated from the \texttt{mad\{c\}.hi/lo} PTX instructions for integer multiply and accumulate on higher and lower 32-bits of the 64-bit product. 
\texttt{IMAD} instructions have a longer issue latency of 4 cycles compared to \texttt{IADD3}'s 2 cycles, and prior works accelerating $FF\_mul$ in elliptic curve signatures convert expensive \texttt{IMAD} instructions to \texttt{IADD3} instructions \cite{gpu-gecc}.
The adaptation of these techniques to ZKP kernels merits further study.

\subsubsection{Field Inverse}\label{subsubsection-ff-ff-inv} The $FF\_inv$ operation, implemented with the binary extended-Euclidean algorithm \cite{gpu-gecc}, is $\sim$100$\times$ slower than $FF\_mul$ due to numerous divide-by-2 operations and branch instructions. 
Given these overheads, $FF\_inv$ is not suitable for GPU acceleration, and MSM implementations do not use elliptic curve forms, like Affine, which require $FF\_inv$ in the $PADD$ and $PDBL$ operations. 
Instead, elliptic curve points are represented using alternate forms like Jacobian and XYZZ, which add additional coordinates (increasing the memory footprint) while replacing $FF\_inv$ operations with other field operations \cite{ec-hyperelliptic-db}. 
The $FF\_op$ mixes for different point representations are shown in Table \ref{table:ec-ops-repr}. $FF\_mul$ and $FF\_sqr$ make up a significant portion of the total number of operations. 

\begin{table}
  \centering
  \setlength{\tabcolsep}{4pt}
   \begin{tabular}{|c|c|c|c|c|c|c|}
    \hline
    \multirow{2}{*}{\textbf{Coordinates}} & \multicolumn{2}{c|}{\textbf{Affine}} & \multicolumn{2}{c|}{\textbf{Jacobian}} & \multicolumn{2}{c|}{\textbf{XYZZ}} \\ 
    & \multicolumn{2}{c|}{$(x, y)$} & \multicolumn{2}{c|}{$(X, Y, Z)$} & \multicolumn{2}{c|}{$(X, Y, ZZ, ZZZ)$} \\ 
    \hline
    \textbf{FF\_op count} & PADD & PDBL & PADD & PDBL & PADD & PDBL \\
    \hline
    \textbf{FF\_add} &  0 &  2 &  1 &  2 &  0 &  1 \\
    \textbf{FF\_sub} &  6 &  4 &  8 &  6 &  6 &  3 \\
    \textbf{FF\_dbl} &  0 &  2 &  5 &  6 &  1 &  3 \\
    \textbf{FF\_mul} &  3 &  2 &  7 &  2 &  8 &  6 \\
    \textbf{FF\_sqr} &  0 &  2 &  4 &  5 &  2 &  3 \\
    \textbf{FF\_inv} &  1 &  1 &  0 &  0 &  0 &  0 \\
    \hline
    \multirow{2}{*}{\textbf{\shortstack{Total\\(FF\_mul/sqr \%)}}}& 10 & 13 & 25 & 21 & 17 & 16 \\
    \cline{2-7}
     & \multicolumn{2}{c|}{43.5} & \multicolumn{2}{c|}{39.1} & \multicolumn{2}{c|}{57.6} \\
    \hline
  \end{tabular}
  \caption{Finite-field operation counts for $PADD$ and $PDBL$ in different coordinate representations.}
  \label{table:ec-ops-repr}
  % \vspace{-0.5cm}
\end{table}

\begin{tcolorbox}[width=\linewidth]
\textbf{\emph{Key Takeaways:}}
  \begin{itemize}[leftmargin=*]
     \item Despite higher per-operation latency than CPU, GPUs leverage their high-throughput architecture to exploit data-level parallelism in ZKP, yielding significant performance gains.
     \item $FF\_mul$ dominates the end-to-end execution time.
     \item $FF\_inv$ is significantly slower than its counterparts, avoiding Affine representation that uses $FF\_inv$ is recommended.
  \end{itemize}
\end{tcolorbox}

%%%%%%%%%%%%%%%%%%%%%%%%%%%%%%%%%%%%%%%%%%%
%%%%%%%%%%%%%%%%%%%%%%%%%%%%%%%%%%%%%%%%%%%
\subsection{Analysis at the Microarchitecture Layer} \label{results-sass-level}

We now explore the performance of the finite-field operations with key GPU microarchitecture metrics collected through NVIDIA Nsight profiling tools. 

\begin{figure}[t]
    \centering
	\includegraphics[width=0.50\textwidth, trim={0.35cm 0cm 0cm 0.3cm}, clip]{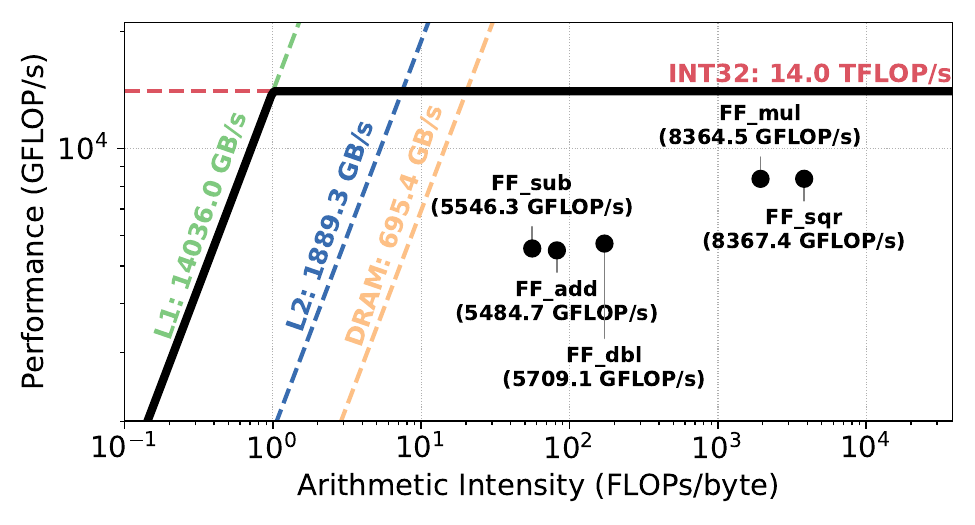}
	\vspace{-0.7cm}
	\caption
	{
	    Roofline analysis of finite-field operations in ZKP.
	}
	\label{figure:roofline}
    %\vspace{-0.5cm}
\end{figure}

\subsubsection{Roofline Analysis.}
Figure \ref{figure:roofline} plots the performance of the finite-field operations within the Roofline envelope of the NVIDIA A40 GPU. 
Throughput of the 32-bit integer execution units determines the compute bound and the bandwidths of the GPU L1, L2, and DRAM determine the respective memory bounds. 
We augment NVIDIA Nsight Compute Roofline Analysis to collect integer instruction metrics as finite-field operations in ZKP rely exclusively on integer instructions.
The GPU performance counters capture the dynamic memory and instruction traffic to calculate Arithmetic Intensity (FLOPs/byte) and Performance (GFLOPs/s).
We assign the 32-bit integer multiply and accumulate instructions (\texttt{IMAD} SASS) a weight of 2 and other integer instructions a weight of 1, consistent with NVIDIA's methodology \cite{nvidia-ampere, ncu-profiling-guide} and prior work \cite{roofline-hierarchy} for floating-point and tensor instructions. 

The figure shows that $FF\_mul$ and $FF\_sqr$ exhibit higher arithmetic intensity than other operations, as they perform more computations per unit of data read.
Looking into the $FF\_op$ performance, we find that $FF\_mul$ and $FF\_sqr$ are able to achieve 60\% of the device's maximum theoretical performance with a majority of \texttt{IMAD} instructions, while $FF\_add$, $FF\_sub$, and $FF\_dbl$ are limited to 40\% of the maximum integer throughput as they primarily rely on \texttt{IADD3} and \texttt{SHF} instructions.
Our analysis shows that the key limiter in the performance of $FF\_op$s is that the GPU schedulers issue new instructions every 3.2 cycles instead of every cycle, and 67.5\% of the cycles see no eligible warps to issue from. 
To understand these warp bottlenecks further, we explore the sources of warp stalls.

\subsubsection{Pipeline Bottlenecks} 

\begin{figure}[t]
    \centering
	\includegraphics[width=0.50\textwidth, trim={0.75cm 0cm 0cm 2cm}, clip]{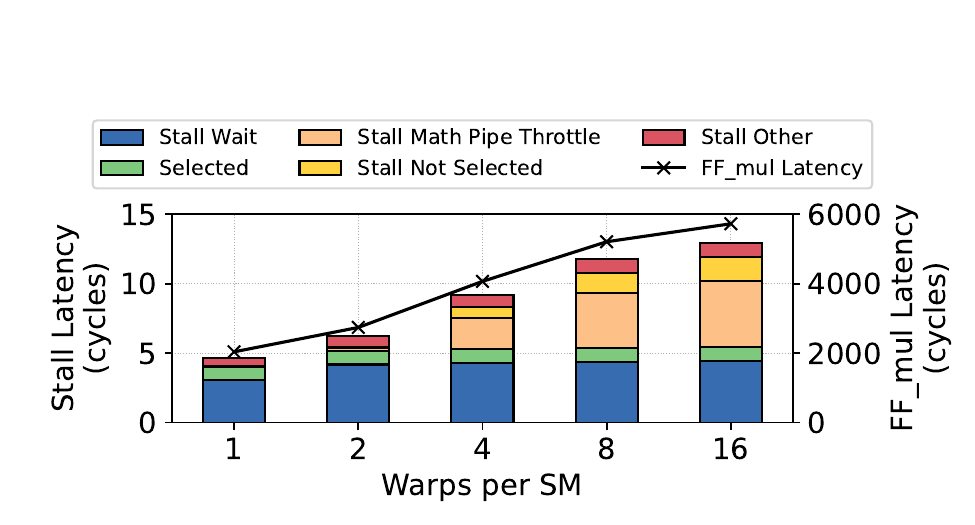}
	\vspace{-0.7cm}
	\caption
	{
	    {Breakdown of warp stalls and latency of $FF\_mul$ operation with varying number of warps per SM.}
	}
	\label{figure:ffmul-warp-stalls}
    \vspace{-0.5cm}
\end{figure}

Figure \ref{figure:ffmul-warp-stalls} shows the average stall latency (in cycles) of the resident warps from different sources.
A higher stall latency implies a worse overall performance for $FF\_mul$. 
The $FF\_mul$ operations, executed with 2 warps per SMSP (representative of MSM configurations), show a warp stall latency of 6.2 cycles. 

The first stall source at $\sim$4 cycles is \emph{Stall Wait}, which denotes a fixed-latency execution dependency. 
$FF\_mul$ executes a series of \texttt{IMAD} instructions, and a new \texttt{IMAD} instruction dependent on the previous one can be issued after the 4 cycle instruction latency, provided there are no other stalls.
The next stall source is \emph{Selected}, which is a 1-cycle latency of issuing a new instruction from the warp. 
This occurs when the warp has been selected by the SMSP scheduler to issue an instruction when its dependencies have been met and the pipeline is available. 

\emph{Stall Math Pipe Throttle} occurs when a specific execution pipeline, in this case the INT32 pipeline, is oversubscribed. 
This is because \textbf{\emph{finite-field operations in ZKPs exclusively use the integer execution units}}. 
Increasing the number of active warps, as suggested in NVIDIA documentation to hide latency \cite{ncu-profiling-guide}, does not improve performance. 
As shown in Figure \ref{figure:ffmul-warp-stalls}, \textbf{\emph{this stall increases as the number of active warps per SM increases, because the warps are still competing for the same limited pipeline}}. 
The other guideline is to utilize additional pipelines by changing the instruction mix. 
This approach merits further exploration, as the floating-point units on the GPU, which have higher throughput, are idle in ZKP kernels. 

The next stall source is \emph{Stall Not Selected}, which specifies that a warp was eligible for being selected but the scheduler picked a different warp to issue from. 
As expected, this stall source increases with additional warps as the pool of ready but bottlenecked warps grows. 

Finally, we study the remaining stall sources from instruction cache misses, branch target computations, memory pipeline throttling, and L1 cache data access. 
These sources combined together are a small fraction of the overall stall latency and they do not increase with additional warps. 
As such, they are reported using \emph{Stall Other} in Figure \ref{figure:ffmul-warp-stalls}. 

We now look at Table~\ref{table:ff-ops-uarch-metrics}, which shows other metrics influencing the performance of $FF\_op$s in MSM and NTT. 

\begin{table}
  \centering
  \setlength{\tabcolsep}{4pt}
   \begin{tabular*}{\linewidth}{@{\extracolsep{\fill}}|c|c|c|c|c|c|}
    \hline
    \textbf{FF\_op / Metric} & \textbf{FF\_add} & \textbf{FF\_sub}
      & \textbf{FF\_dbl} & \textbf{FF\_mul}
      & \textbf{FF\_sqr} \\
    \hline
    \textbf{\shortstack{Branch \\Efficiency (\%)}} & 52.5 & 56.2 & 77.5 & 84.0 & 96.9 \\
    \hline
    \textbf{\shortstack{Achieved\\Occupancy (\%)}} & \multicolumn{5}{c|}{25.0} \\
    \hline
    \textbf{\shortstack{Dominant SASS\\Instruction (\%)}} & \texttt{IADD3} & \texttt{IADD3} & \texttt{SHF} & \texttt{IMAD} & \texttt{IMAD} \\
    \hline
    \textbf{\shortstack{Pipeline\\Bottleneck}} & Integer & Integer & Integer & Integer & Integer \\
    \hline
    %\hline
    %\textbf{\shortstack{Uncoalesced \\Sectors (\%)}} & \multicolumn{5}{c|}{85} \\
    %\hline
  \end{tabular*}
  \caption{GPU microarchitecture metrics for $FF\_op$s.}
  \label{table:ff-ops-uarch-metrics}
  \vspace{-0.5cm}
\end{table}

\subsubsection{Branch Efficiency} 

This metric denotes the proportion of branch targets where all (active) threads of a warp select the same target. 
In other words, a branch efficiency of 100\% implies no thread divergence in the warp. 
$FF\_add$ and $FF\_sub$ exhibit branch efficiencies of 52.5\% and 56.2\% respectively, stemming from the sequential comparison between the corresponding limbs of the result and the field modulus/$zero$.
These divergences causes a 2.4$\times$ increase in execution cycles (72 to 244 as discussed in \S\ref{results-ff-add-sub-dbl}). 

$FF\_dbl$ has a higher branch efficiency of 77.5\%. 
When doubling is performed with $FF\_add$, the efficiency of $FF\_add$ also jumps up to 77.5\%. 
The inputs for the G1 MSM kernel which are processed by the finite-field operations are uniformly random \cite{gpu-cuzk, gpu-ymc}. 
Addition with a random field element is more likely to result in one of the limbs ending up outside the corresponding field modulus limb for any one of the 32 threads in a warp.

$FF\_mul$ and $FF\_sqr$ have much higher branch efficiencies of 84.0\% and 96.9\% respectively.
The multiplication algorithm performs a cross-product of all the limbs, addition of higher and lower 32-bit parts of 64-bit products, followed by the reduction operations\cite{gpu-gecc, opt-modmul}. 
The branch efficiency is high because most products require a final reduction operation, with the only difference being in which of the limbs is outside the modulus range. 
The branch divergence is only responsible for 3.8\% of the total cycles in $FF\_mul$ and $FF\_sqr$ compared to 70.5\% for $FF\_add$ and $FF\_sub$. 
\textbf{\emph{Branch efficiency is critical for optimizing performance, as this metric is less than 50\% in MSM implementations.}}

\subsubsection{Achieved Occupancy} 

Occupancy refers to the number of blocks resident on each GPU SM.  
Theoretical Occupancy is determined by the GPU's Compute Capability (CC), per-thread register usage and per-block shared-memory usage, while Achieved Occupancy is determined by launch configuration of $<<<$blocks, threads$>>>$.
While a higher occupancy can hide latencies and enable better GPU utilization, it is not necessary to achieve optimal performance \cite{volkov-dissertation}. 
\emph{bellperson} employs a launch configuration of $<<<$168, 128$>>>$, while \emph{sppark} and \emph{ymc} use launch configurations of $<<<$84, 128$>>>$ for MSM kernels on NVIDIA A40.
\emph{ymc} hides memory latencies using asynchronous memory transfers between the GPU memories and caches (introduced in the Ampere \cite{nvidia-ampere} microarchitecture). 
MSM kernels additionally exhibit high register usage: \emph{bellperson}, \emph{sppark}, and \emph{ymc} require up to 228, 216, and 244 registers per thread. 
A large number of live registers are required to perform $FF\_mul$ operations on 4 12-limb (up to 384-bit integers) coordinates in the XYZZ representation. 
NTT has a lower live register count of 56, since (1) each scalar is a single 8-limb element and (2) the dependence chain of $FF\_op$s is much shorter in NTT butterfly operations than MSM $PADD$s.

\begin{tcolorbox}[width=\linewidth]
\textbf{\emph{Key Takeaways:}}
  \begin{itemize}[leftmargin=*]
    \item With memory latencies hidden, the performance is bottlenecked by INT32 cores.
    \item Adding additional threads may increase stalls and negatively impact performance.
    \item Branch efficiency is critical for ZKP performance.
  \end{itemize}
\end{tcolorbox}

%%%%%%%%%%%%%%%%%%%%%%%%%%%%%%%%%%%%%%%%%%%
%%%%%%%%%%%%%%%%%%%%%%%%%%%%%%%%%%%%%%%%%%%
\subsection{FF\_mul performance across GPU generations}\label{results-diff-gpus}

Recent GPU advancements have been primarily motivated by AI models like LLMs, and this section explores how these innovations may benefit ZKP workloads.
We study the performance of $FF\_mul$ over multiple generations of GPU architectures to understand speedup sources and discover future performance improvement opportunities for ZKPs.

\begin{figure}[t]
    \centering
    \begin{subfigure}{\linewidth}
        \centering
        \includegraphics[width=\linewidth, trim={0cm 0cm 0cm 0.5cm}, clip]{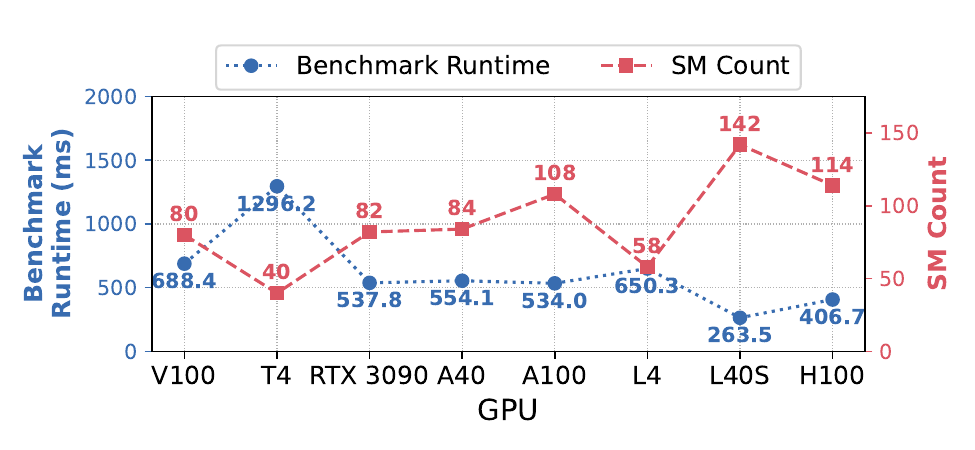}
        \vspace{-0.7cm}
        %\caption{SM Count and $FF\_mul$ benchmark runtime}
        \caption{$FF\_mul$ benchmark runtime and SM Count.}
        \label{figure:ffmul-diff-gpus-perf-sm}
    \end{subfigure}
    \begin{subfigure}{\linewidth}
        \centering
        \includegraphics[width=\linewidth, trim={0cm 0cm 0cm 0.5cm}, clip]{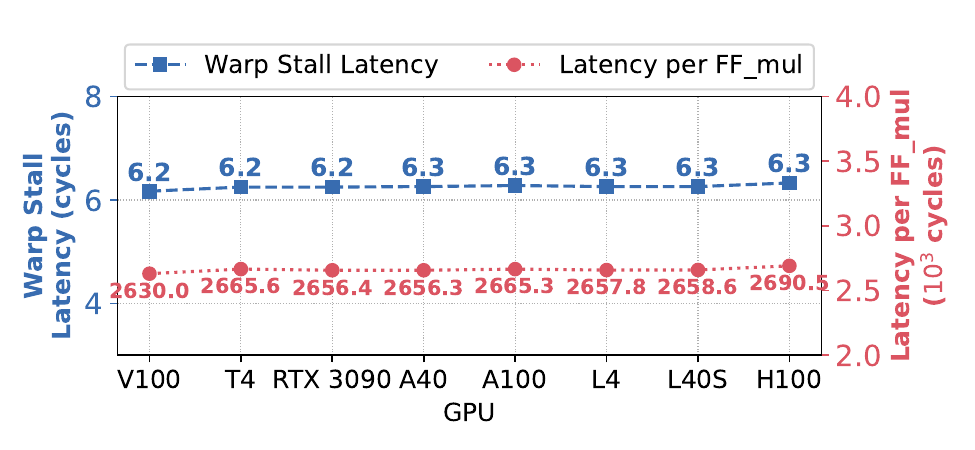}
        \vspace{-0.7cm}
        \caption{Warp stall latency and latency of each $FF\_mul$.}
        \label{figure:ffmul-diff-gpus-wlpi}
    \end{subfigure}
	\caption
	{
	    {$FF\_mul$ performance across GPU generations.}
	}
	\label{figure:ffmul-diff-gpus}
    \vspace{-0.7cm}
\end{figure}

Since the benchmark scales well across the number of available SMs, we observe that \textbf{\emph{the runtime is inversely proportional to the number of SMs}}. 
Figure~\ref{figure:ffmul-diff-gpus-perf-sm} shows that the NVIDIA L40S (CC 8.9), with 24.6\% more SMs, is 1.5$\times$ faster than NVIDIA H100 (CC 9.0). 

For a deeper analysis of the $FF\_mul$ performance profile, we analyze the warp stall sources and the latency per $FF\_mul$ and plot the results in Figure \ref{figure:ffmul-diff-gpus-wlpi}.
We see that the warp stall latency is consistent across the 8 GPUs evaluated with an average value of 6.26. 
As discussed in \S\ref{results-sass-level}, this stall latency encapsulates the effects of different microarchitectural features, and we see a similar stall latency breakdown across GPU generations.
Consequently, \textbf{\emph{the latency per $FF\_mul$ in cycles is also constant at 2660.06 cycles on average}}. 

Put together, these results reveal that the performance scaling of existing well-optimized NTT \cite{gpu-cuzk} and MSM \cite{gpu-sppark, gpu-cuzk, gpu-ymc} implementations is primarily driven by additional SM units available on the GPU, with the per SM performance being more or less constant.
While newer GPU generations offer improved memory bandwidth, existing MSM implementations hide the memory latency well using the asynchronous data transfers between the global memory and caches/shared memory. 
Metrics determining the performance at the microarchitecture level, such as registers/thread, warp size, 32-bit \texttt{IMAD} throughput, and the number of INT32 pipelines, have been constant across several generations of NVIDIA architectures \cite{nvidia-programming-guide}. 
Other architectural improvements focus on tensor cores, which are unused in $FF\_op$s.

\subsubsection{Additional performance scaling opportunities}

Successive generations of NVIDIA GPU architectures have focused on improving the GPU memory bandwidth and cache/shared-memory capacities along with increasing the available GPU memory. 
Since the per-SM INT32 performance remains the same, we can employ optimizations with fewer $FF\_mul$ operations at the cost of additional memory usage.
We briefly discuss two optimizations.

\paragraph{\textbf{Reducing windows with precomputation.}}
As discussed in \S\ref{subsection:background-msm}, the Bucket Reduction step uses the Sum-of-Sums algorithm to compute the sum of each window using $2 * 2^c$ $PADD$s per window. 
For a representative window size of $c = 23$ bits, a 253-bit scalar requires $w = 11$ windows, and each window thus requires 16.7 million $PADD$s. 
To reduce the number of windows, we can precompute $2^{q*c} * P_i$ for each point $P_i$.
Then, instead of adding point $P_j$ to window $q = 3$, we can add $2^{3*c} * P_j$ to window 0.
This optimization is especially useful when processing a batch of MSMs where the points $P_i$ are fixed \cite{gpu-gzkp, gpu-ymc}.

\begin{figure}[t]
    \centering
	\includegraphics[width=\linewidth, trim={0cm 0cm 0cm 0cm}, clip]{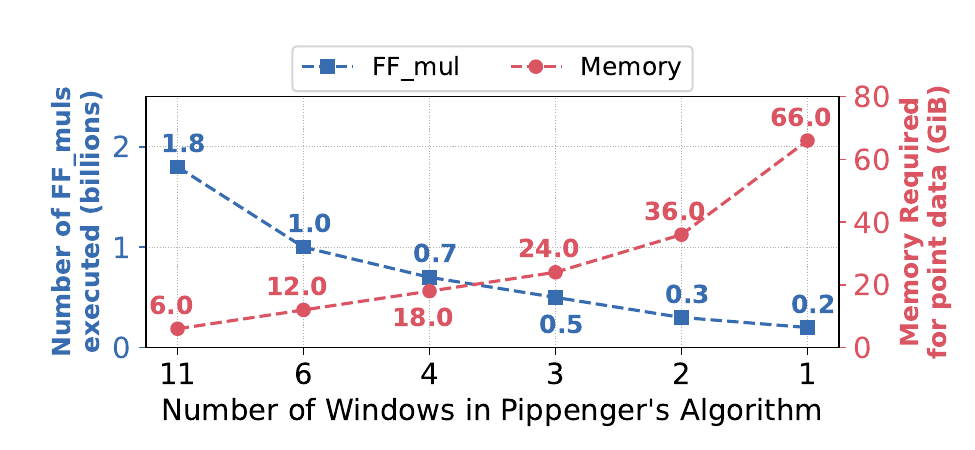}
    \vspace{-0.7cm}
	\caption
	{
	    {Number of $FF\_mul$s and memory usage for storing elliptic curve points across Pippenger windows.}
	}
	\label{figure:precomp-ops-mem}
    %\vspace{-0.7cm}
\end{figure}

For scale $n = 2^{26}$, each set of points (represented initially in Affine form) requires 6 GiB of memory.
Given a window size $c = 23$ and 10 $FF\_mul$ operations per $PADD$, Figure \ref{figure:precomp-ops-mem} plots the number of $FF\_mul$ operations required for Bucket Reduction as we decrease the number of windows by precomputing additional points.
The figure also shows the storage required (in GiBs) to store the precomputed points on the GPU memory. 
\textbf{\emph{Reducing the number of windows through precomputation can significantly reduce the number of $FF\_mul$s, provided enough device memory is available. }}
For example, the MSM can be executed with $w=4$ windows on the 24 GB NVIDIA L40, with $w=2$ windows on the 48 GB NVIDIA A40, and $w=1$ window on the 80 GB NVIDIA A100 and H100 GPUs. 

\paragraph{Representing points in Affine form}
The Montgomery Trick for Batched Inversion \cite{opt-batched-affine} replaces $N$ $FF\_inv$s with $1$ $FF\_inv$ and $3N$ $FF\_mul$ for Affine forms.
To calculate the inverses of $N$ elements, this approach multiplies the $N$ elements together and performs a single inverse of the final product. 
During these multiplications, the partial products generated at each step of the multiplication are stored. 
Since the multiplications are done in a finite field, the partial and final products are constant-sized field elements. 
These partial products are then multiplied individually with the inverse of the final product to extract the $N$ inverses.

For $n = 2^{26}$ elliptic curve point additions, the Affine representation and the Montgomery Trick can reduce the number of $FF\_mul$ operations by 3.3$\times$ and 3.6$\times$ for the XYZZ and Jacobian representations, provided that the cost of the $FF\_inv$ can be amortized over a large enough batch size. 
This approach produces a lot of intermediate data in terms of the partial products and their inverses, which exceed the L1/L2 cache sizes of the GPU, leading to expensive global memory accesses. 
For example, a batch size of $2^{20}$ for a large scale MSM would require an additional 300 MB for storing the intermediate products and the inverses, which exceeds the 40 MB and 50 MB L2 caches of the NVIDIA A100 and NVIDIA H100 respectively. 
%(2*((2^22)-1)+2^22)*2*12*32 bits to mb

Prior work \cite{gpu-gecc} implements the Montgomery Trick for a throughput-focused elliptic curve arithmetic library by employing Gather-Apply-Scatter techniques over the warps and fine-grained control of the memory hierarchy. 
The application of these techniques to the MSM algorithm by leveraging the increasing memory sizes and bandwidths on modern GPUs requires further study. 

\begin{tcolorbox}[width=\linewidth]
\textbf{\emph{Key Takeaways:}}
  \begin{itemize}[leftmargin=*]
     \item ZKP performance is driven by the number of SMs on the GPU.
     \item Per-SM performance of $FF\_mul$ is constant across several GPU generations.
     \item Future optimizations should utilize the growing memory capacities and bandwidths in GPUs.
  \end{itemize}
\end{tcolorbox}

\section{Next Steps in the ZKP Ecosystem}\label{section:discussion}
Based on our quantitative analyses, we provide several recommendations for future software and hardware developments in the GPU-accelerated ZKP ecosystem. 

\subsection{For ZKP Application Developers}
Optimizing proof generation on GPUs requires selecting appropriate kernel implementations based on the application circuit size, and relevant recommendations are provided in Table \ref{table:msm-ntt-speedup}.
However, these implementations offer limited interoperability with each other and with end-to-end ZKP frameworks (like \emph{arkworks} \cite{cpu-arkworks}, \emph{libsnark} \cite{cpu-libsnark}, and \emph{xjsnark} \cite{cpu-xjsnark}), which convert the application code into inputs for the $Prover$.
End-user ZKP applications \cite{app-zkmb, app-filecoin, app-zkdl} thus utilize CPU-based or slower GPU-based $Prover$s and miss out on orders of magnitude of speedups.
Constructing a high-performance $Prover$ would thus require manual effort to integrate the required components.
Accelerated implementations often feature hand-tuned optimizations for underlying elliptic curves and finite fields with custom algorithms and PTX routines, and they may target a specific GPU architecture or chip, further hindering interoperability. 
Additionally, the proof generation design space spans several parameters like (1) the choice of the framework to generate constraints, (2) the elliptic curves and finite fields to encode inputs, (3) kernel-specific optimizations like precomputed inputs, all driven by (4) hardware parameters like available GPU compute units and global and shared memories.
These numerous tunable parameters are currently manually picked for each application, motivating the development of autotuning tools which can optimally adapt an application to a Zero-Knowledge Proof on the target GPU at runtime. 

\subsection{For GPU Kernel Programmers}
State-of-the-art MSM kernels \cite{gpu-distmsm, gpu-sppark, gpu-ymc, opt-zprize} offer speedups over CPUs up to 800$\times$, while ZKP-compatible NTT kernels are limited to 50$\times$ and constrain end-to-end speedup (Figure \ref{figure:proof-gpu-speedup}).
NTT should therefore be a key acceleration target moving forward. 
Prior work \cite{ntt-f1, ntt-craterlake, ntt-rpu, ntt-bts, ntt-gjcc, ntt-kjpa} can be adapted to ZKPs while accounting for larger bitwidth requirements and utilizing optimized finite-field routines from MSM libraries \cite{gpu-sppark, gpu-ymc}.
Optimized kernels are often tailored to a specific GPU and may under-utilize the resources of newer generations of GPUs.
For example, NVIDIA H100 with 80 GB of memory can support more precomputed windows in the \emph{ymc} MSM implementation to extract further speedup (\S\ref{results-diff-gpus}). 
Future implementations should therefore ensure maximum utilization of available hardware resources.  
Moving forward, accelerated kernels should emphasize interoperability with end-to-end ZKP frameworks to enable wider adoption.
\emph{arkworks} \cite{cpu-arkworks} is a viable target given its support for a variety of proof systems, elliptic curves, and finite fields as well as adoption in industry-driven acceleration efforts \cite{opt-zprize}. 
Finally, alternate proof systems like Orion \cite{zkp-orion}, STARK \cite{zkp-stark}, and Aurora \cite{zkp-aurora} can be accelerated with GPU implementations of primitives like vector operations, hashing, and Sum-Check \cite{zkp-icicle-v2}. 

\subsection{For GPU Architecture Designers}
Our analysis in \S\ref{results-diff-gpus} shows that the performance of 32-bit integer pipelines has remained constant across several generations of GPU architectures.
Since optimized MSM implementations like \cite{gpu-sppark, gpu-ymc, gpu-distmsm} utilize the available memory bandwidth by overlapping compute with data movement between different levels of the GPU memory hierarchy, $Prover$ performance improvements primarily stem from adding Streaming Multiprocessors. 
The SM architectural improvements have been restricted to 32-bit floating-point pipelines, which double the instruction throughput from the Ampere generation onward by utilizing the integer pipelines, and Tensor Cores, which offer generational performance improvements for low precision (4-16 bit) integer and floating-point types. 
Adapting these architectural improvements for 32-bit integer computations (specifically the \texttt{IMAD} instruction), supporting higher-precision integer computations (similar to 64-bit floating-point instructions), concurrently utilizing integer and floating-point units \cite{gpu-gecc}, and supporting higher-precision arithmetic in Tensor Cores can further scale ZKP performance on GPUs. 

\section{Related Work}\label{section:related-work}
ZKPs are powerful cryptographic primitives with applications in private and verifiable computing \cite{app-filecoin, app-zcash, app-celo, app-loopring, app-ing, app-zkmb, app-zombie, app-greco, app-verif-ml, app-nope, app-darkforest, app-zkdl, app-hermez, app-semaphore}.
Groth16 \cite{zkp-groth} is a popular proof system due to its compact proof sizes and constant verification time, and its MSM and NTT kernels have been accelerated on CPUs \cite{cpu-arkworks, cpu-dizk, cpu-edmsm}, GPUs \cite{gpu-bellperson, gpu-cuzk, gpu-distmsm, gpu-elasticmsm, gpu-gzkp, gpu-icicle, gpu-sppark, gpu-ymc, opt-zprize}, FPGAs \cite{fpga-bstmsm, fpga-falic, fpga-hardcaml, fpga-msmac, fpga-pipemsm}, and ASICs \cite{asic-pipezk, asic-szkp, asic-priormsm}. 
Additional efforts \cite{asic-nocap, gpu-batchzk, gpu-zkpog, asic-zkspeed} target alternate proof systems which offer improved $Prover$ performance but at the cost of $Verifier$ performance and increased proof sizes. Proofs generated by these protocols can be combined with Groth16 proofs to for small proof sizes and constant $Verifier$ latency. 
NTT acceleration efforts have primarily been driven by HE applications \cite{ntt-hexl, ntt-f1, ntt-craterlake, ntt-rpu, ntt-bts, ntt-gjcc, ntt-kjpa, ntt-wlhm} and merit further study for ZKPs.
\cite{zkp-cpu-perf} presents a top-down analysis of CPU implementations of the Groth16 protocol, focusing on higher-level stages (Compile, Setup, Witness, Proving, and Verifying) of ZK-SNARKs. 
In contrast, ZKProphet presents a detailed analysis of the Proving step on GPUs, the primary hardware platform for proof generation.

To the best of our knowledge, ZKProphet is the first work performing detailed performance analysis of ZKP proof generation workloads on GPUs.

\section{Conclusion}\label{section:conclusion}
We present ZKProphet, a detailed performance characterization of end-to-end proof generation on GPUs.
We find that state-of-the-art libraries significantly optimize Multi-Scalar Multiplication, shifting the performance bottleneck to Number-Theoretic Transform. 
The fastest proof-generation framework at a particular application size typically comprises of kernels from different libraries, which offer limited compatibility each other and require manual integration efforts.
Through detailed microarchitectural studies on a diverse set of GPUs, we identify that performance scaling of ZKPs is limited by the GPUs' integer execution units, and critical GPU resources are often underutilized.
End-to-end proof generation on GPUs can be further optimized by tailoring implementations to application parameters and available GPU resources.

\bibliographystyle{IEEEtranS}
\bibliography{99_zkp-refs}

% Generated by IEEEtranS.bst, version: 1.12 (2007/01/11)
\begin{thebibliography}{10}
\providecommand{\url}[1]{#1}
\csname url@samestyle\endcsname
\providecommand{\newblock}{\relax}
\providecommand{\bibinfo}[2]{#2}
\providecommand{\BIBentrySTDinterwordspacing}{\spaceskip=0pt\relax}
\providecommand{\BIBentryALTinterwordstretchfactor}{4}
\providecommand{\BIBentryALTinterwordspacing}{\spaceskip=\fontdimen2\font plus
\BIBentryALTinterwordstretchfactor\fontdimen3\font minus \fontdimen4\font\relax}
\providecommand{\BIBforeignlanguage}[2]{{%
\expandafter\ifx\csname l@#1\endcsname\relax
\typeout{** WARNING: IEEEtranS.bst: No hyphenation pattern has been}%
\typeout{** loaded for the language `#1'. Using the pattern for}%
\typeout{** the default language instead.}%
\else
\language=\csname l@#1\endcsname
\fi
#2}}
\providecommand{\BIBdecl}{\relax}
\BIBdecl

\bibitem{zkp-ligero}
\BIBentryALTinterwordspacing
S.~Ames, C.~Hazay, Y.~Ishai, and M.~Venkitasubramaniam, ``Ligero: Lightweight sublinear arguments without a trusted setup,'' Cryptology {ePrint} Archive, Paper 2022/1608, 2022. [Online]. Available: \url{https://eprint.iacr.org/2022/1608}
\BIBentrySTDinterwordspacing

\bibitem{app-darkforest}
\BIBentryALTinterwordspacing
arkforest\_eth, ``Dark forest, the world's first decentralized real-time strategy game,'' 2020. [Online]. Available: \url{https://zkga.me/}
\BIBentrySTDinterwordspacing

\bibitem{cpu-arkworks}
\BIBentryALTinterwordspacing
arkworks contributors, ``\texttt{arkworks} zksnark ecosystem,'' 2022. [Online]. Available: \url{https://arkworks.rs}
\BIBentrySTDinterwordspacing

\bibitem{zkp-stark}
\BIBentryALTinterwordspacing
E.~Ben-Sasson, I.~Bentov, Y.~Horesh, and M.~Riabzev, ``Scalable, transparent, and post-quantum secure computational integrity,'' Cryptology {ePrint} Archive, Paper 2018/046, 2018. [Online]. Available: \url{https://eprint.iacr.org/2018/046}
\BIBentrySTDinterwordspacing

\bibitem{zkp-aurora}
E.~Ben-Sasson, A.~Chiesa, M.~Riabzev, N.~Spooner, M.~Virza, and N.~P. Ward, ``Aurora: Transparent succinct arguments for r1cs,'' in \emph{Advances in Cryptology – EUROCRYPT 2019: 38th Annual International Conference on the Theory and Applications of Cryptographic Techniques, Darmstadt, Germany, May 19–23, 2019, Proceedings, Part I}.\hskip 1em plus 0.5em minus 0.4em\relax Springer-Verlag, 2019, p. 103–128.

\bibitem{ec-hyperelliptic-db}
\BIBentryALTinterwordspacing
D.~J. Bernstein and T.~Lange, ``The explicit-formulas database.'' [Online]. Available: \url{https://www.hyperelliptic.org/EFD/index.html}
\BIBentrySTDinterwordspacing

\bibitem{ntt-hexl}
\BIBentryALTinterwordspacing
F.~Boemer, S.~Kim, G.~Seifu, F.~D.~M. de~Souza, and V.~Gopal, ``Intel hexl: Accelerating homomorphic encryption with intel avx512-ifma52,'' 2021. [Online]. Available: \url{https://arxiv.org/abs/2103.16400}
\BIBentrySTDinterwordspacing

\bibitem{zkp-recursive-halo-inf}
\BIBentryALTinterwordspacing
D.~Boneh, J.~Drake, B.~Fisch, and A.~Gabizon, ``Halo infinite: Recursive zk-{SNARKs} from any additive polynomial commitment scheme,'' Cryptology {ePrint} Archive, Paper 2020/1536, 2020. [Online]. Available: \url{https://eprint.iacr.org/2020/1536}
\BIBentrySTDinterwordspacing

\bibitem{app-greco}
\BIBentryALTinterwordspacing
E.~Bottazzi, ``Greco: Fast zero-knowledge proofs for valid {FHE} {RLWE} ciphertexts formation,'' Cryptology {ePrint} Archive, Paper 2024/594, 2024. [Online]. Available: \url{https://eprint.iacr.org/2024/594}
\BIBentrySTDinterwordspacing

\bibitem{zkp-bulletproofs}
B.~Bünz, J.~Bootle, D.~Boneh, A.~Poelstra, P.~Wuille, and G.~Maxwell, ``Bulletproofs: Short proofs for confidential transactions and more,'' in \emph{2018 IEEE Symposium on Security and Privacy (SP)}, 2018, pp. 315--334.

\bibitem{zkp-marlin}
\BIBentryALTinterwordspacing
A.~Chiesa, Y.~Hu, M.~Maller, P.~Mishra, P.~Vesely, and N.~Ward, ``Marlin: Preprocessing {zkSNARKs} with universal and updatable {SRS},'' Cryptology {ePrint} Archive, Paper 2019/1047, 2019. [Online]. Available: \url{https://eprint.iacr.org/2019/1047}
\BIBentrySTDinterwordspacing

\bibitem{cpu-gnark}
\BIBentryALTinterwordspacing
Consensys, ``gnark zk-snark library,'' Feb 2020. [Online]. Available: \url{https://github.com/Consensys/gnark}
\BIBentrySTDinterwordspacing

\bibitem{ntt-cooley-tukey}
J.~Cooley and J.~Tukey, ``An algorithm for the machine calculation of complex fourier series,'' \emph{Mathematics of Computation}, vol.~19, pp. 297--301, 1965.

\bibitem{asic-zkspeed}
\BIBentryALTinterwordspacing
A.~Daftardar, J.~Mo, J.~Ah-kiow, B.~B\"{u}nz, R.~Karri, S.~Garg, and B.~Reagen, ``Need for zkspeed: Accelerating hyperplonk for zero-knowledge proofs,'' in \emph{Proceedings of the 52nd Annual International Symposium on Computer Architecture}, ser. ISCA '25.\hskip 1em plus 0.5em minus 0.4em\relax New York, NY, USA: Association for Computing Machinery, 2025, p. 1986–2001. [Online]. Available: \url{https://doi.org/10.1145/3695053.3731021}
\BIBentrySTDinterwordspacing

\bibitem{asic-szkp}
\BIBentryALTinterwordspacing
A.~Daftardar, B.~Reagen, and S.~Garg, ``Szkp: A scalable accelerator architecture for zero-knowledge proofs,'' in \emph{Proceedings of the 2024 International Conference on Parallel Architectures and Compilation Techniques}, ser. PACT ’24.\hskip 1em plus 0.5em minus 0.4em\relax ACM, Oct. 2024, p. 271–283. [Online]. Available: \url{http://dx.doi.org/10.1145/3656019.3676898}
\BIBentrySTDinterwordspacing

\bibitem{app-nope}
\BIBentryALTinterwordspacing
Z.~DeStefano, J.~J. Ma, J.~Bonneau, and M.~Walfish, ``Nope: Strengthening domain authentication with succinct proofs,'' in \emph{Proceedings of the ACM SIGOPS 30th Symposium on Operating Systems Principles}, ser. SOSP '24.\hskip 1em plus 0.5em minus 0.4em\relax New York, NY, USA: Association for Computing Machinery, 2024, p. 673–692. [Online]. Available: \url{https://doi.org/10.1145/3694715.3695962}
\BIBentrySTDinterwordspacing

\bibitem{opt-modmul}
N.~Emmart, J.~Luitjens, C.~Weems, and C.~Woolley, ``Optimizing modular multiplication for nvidia's maxwell gpus,'' in \emph{2016 IEEE 23nd Symposium on Computer Arithmetic (ARITH)}, 2016, pp. 47--54.

\bibitem{app-semaphore}
\BIBentryALTinterwordspacing
P.~.~S. Explorations, ``Semaphore: A zero-knowledge protocol for anonymous interactions,'' 2019. [Online]. Available: \url{https://semaphore.pse.dev/}
\BIBentrySTDinterwordspacing

\bibitem{gpu-bellperson}
\BIBentryALTinterwordspacing
Filecoin, ``bellperson: zk-snark library,'' Online. [Online]. Available: \url{https://github.com/filecoin-project/bellperson}
\BIBentrySTDinterwordspacing

\bibitem{app-filecoin}
\BIBentryALTinterwordspacing
------, ``Filecoin: A decentralized storage network for the world's information,'' 2014. [Online]. Available: \url{https://filecoin.io/}
\BIBentrySTDinterwordspacing

\bibitem{app-celo}
\BIBentryALTinterwordspacing
C.~Foundation, ``Celo blockchain,'' 2017. [Online]. Available: \url{https://celo.org/}
\BIBentrySTDinterwordspacing

\bibitem{opt-batched-affine}
\BIBentryALTinterwordspacing
A.~Gabizon and Z.~J. Williamson, ``Proposal: The turbo-plonk program syntax for specifying snark programs,'' in \emph{3rd ZKProof Workshop}, 2020. [Online]. Available: \url{https://docs.zkproof.org/pages/standards/accepted-workshop3/proposal-turbo\_plonk.pdf}
\BIBentrySTDinterwordspacing

\bibitem{zkp-plonk}
\BIBentryALTinterwordspacing
A.~Gabizon, Z.~J. Williamson, and O.~Ciobotaru, ``{PLONK}: Permutations over lagrange-bases for oecumenical noninteractive arguments of knowledge,'' Cryptology {ePrint} Archive, Paper 2019/953, 2019. [Online]. Available: \url{https://eprint.iacr.org/2019/953}
\BIBentrySTDinterwordspacing

\bibitem{zkp-groth}
\BIBentryALTinterwordspacing
J.~Groth, ``On the size of pairing-based non-interactive arguments,'' Cryptology ePrint Archive, Paper 2016/260, 2016, \url{https://eprint.iacr.org/2016/260}. [Online]. Available: \url{https://eprint.iacr.org/2016/260}
\BIBentrySTDinterwordspacing

\bibitem{app-zkmb}
\BIBentryALTinterwordspacing
P.~Grubbs, A.~Arun, Y.~Zhang, J.~Bonneau, and M.~Walfish, ``{Zero-Knowledge} middleboxes,'' in \emph{31st USENIX Security Symposium (USENIX Security 22)}.\hskip 1em plus 0.5em minus 0.4em\relax Boston, MA: USENIX Association, Aug. 2022, pp. 4255--4272. [Online]. Available: \url{https://www.usenix.org/conference/usenixsecurity22/presentation/grubbs}
\BIBentrySTDinterwordspacing

\bibitem{ntt-gjcc}
N.~Gupta, A.~Jati, A.~K. Chauhan, and A.~Chattopadhyay, ``Pqc acceleration using gpus: Frodokem, newhope, and kyber,'' \emph{IEEE Transactions on Parallel and Distributed Systems}, vol.~32, no.~3, pp. 575--586, 2021.

\bibitem{cpu-edmsm}
\BIBentryALTinterwordspacing
Y.~E. Housni and G.~Botrel, ``{EdMSM}: Multi-scalar-multiplication for {SNARKs} and faster montgomery multiplication,'' Cryptology {ePrint} Archive, Paper 2022/1400, 2022. [Online]. Available: \url{https://eprint.iacr.org/2022/1400}
\BIBentrySTDinterwordspacing

\bibitem{cpu-circom}
\BIBentryALTinterwordspacing
iden3, ``Circom: Circuit compiler for zero- knowledge proofs,'' 2020. [Online]. Available: \url{https://github.com/iden3/circom}
\BIBentrySTDinterwordspacing

\bibitem{zkp-icicle-v2}
\BIBentryALTinterwordspacing
K.~Inbasekar, Y.~Shekel, and M.~Asa, ``{ICICLE} v2: Polynomial {API} for coding {ZK} provers to run on specialized hardware,'' Cryptology {ePrint} Archive, Paper 2024/973, 2024. [Online]. Available: \url{https://eprint.iacr.org/2024/973}
\BIBentrySTDinterwordspacing

\bibitem{gpu-icicle}
\BIBentryALTinterwordspacing
Ingonyama, ``Icicle: A hardware acceleration library for compute intensive cryptography,'' Online. [Online]. Available: \url{https://github.com/ingonyama-zk/icicle}
\BIBentrySTDinterwordspacing

\bibitem{gpu-distmsm}
\BIBentryALTinterwordspacing
Z.~Ji, Z.~Zhang, J.~Xu, and L.~Ju, ``Accelerating multi-scalar multiplication for efficient zero knowledge proofs with multi-gpu systems,'' in \emph{Proceedings of the 29th ACM International Conference on Architectural Support for Programming Languages and Operating Systems, Volume 3}, ser. ASPLOS '24.\hskip 1em plus 0.5em minus 0.4em\relax New York, NY, USA: Association for Computing Machinery, 2024, p. 57–70. [Online]. Available: \url{https://doi.org/10.1145/3620666.3651364}
\BIBentrySTDinterwordspacing

\bibitem{app-verif-ml}
\BIBentryALTinterwordspacing
D.~Kang, T.~Hashimoto, I.~Stoica, and Y.~Sun, ``Scaling up trustless dnn inference with zero-knowledge proofs,'' 2022. [Online]. Available: \url{https://arxiv.org/abs/2210.08674}
\BIBentrySTDinterwordspacing

\bibitem{ntt-kjpa}
\BIBentryALTinterwordspacing
S.~Kim, W.~Jung, J.~Park, and J.~H. Ahn, ``{ Accelerating Number Theoretic Transformations for Bootstrappable Homomorphic Encryption on GPUs },'' in \emph{2020 IEEE International Symposium on Workload Characterization (IISWC)}.\hskip 1em plus 0.5em minus 0.4em\relax Los Alamitos, CA, USA: IEEE Computer Society, Oct. 2020, pp. 264--275. [Online]. Available: \url{https://doi.ieeecomputersociety.org/10.1109/IISWC50251.2020.00033}
\BIBentrySTDinterwordspacing

\bibitem{ntt-bts}
\BIBentryALTinterwordspacing
S.~Kim, J.~Kim, M.~J. Kim, W.~Jung, J.~Kim, M.~Rhu, and J.~H. Ahn, ``Bts: an accelerator for bootstrappable fully homomorphic encryption,'' in \emph{Proceedings of the 49th Annual International Symposium on Computer Architecture}, ser. ISCA ’22.\hskip 1em plus 0.5em minus 0.4em\relax ACM, Jun. 2022, p. 711–725. [Online]. Available: \url{http://dx.doi.org/10.1145/3470496.3527415}
\BIBentrySTDinterwordspacing

\bibitem{cpu-jsnark}
\BIBentryALTinterwordspacing
A.~Kosba, ``jsnark: a java library for building circuits for preprocessing zk-snarks.'' 2019. [Online]. Available: \url{https://github.com/akosba/jsnark}
\BIBentrySTDinterwordspacing

\bibitem{cpu-xjsnark}
\BIBentryALTinterwordspacing
A.~Kosba, C.~Papamanthou, and E.~Shi, ``{ xJsnark: A Framework for Efficient Verifiable Computation },'' in \emph{2018 IEEE Symposium on Security and Privacy (SP)}.\hskip 1em plus 0.5em minus 0.4em\relax Los Alamitos, CA, USA: IEEE Computer Society, May 2018, pp. 944--961. [Online]. Available: \url{https://doi.ieeecomputersociety.org/10.1109/SP.2018.00018}
\BIBentrySTDinterwordspacing

\bibitem{gpu-zkpog}
\BIBentryALTinterwordspacing
M.~Li, Y.~Yu, B.~Wang, X.~Fan, and S.~Deng, ``{ZKPoG}: Accelerating {WitGen}-incorporated end-to-end zero-knowledge proof on {GPU},'' Cryptology {ePrint} Archive, Paper 2025/765, 2025. [Online]. Available: \url{https://eprint.iacr.org/2025/765}
\BIBentrySTDinterwordspacing

\bibitem{cpu-libsnark}
\BIBentryALTinterwordspacing
libsnark contributors, ``libsnark: a c++ library for zksnark proofs,'' 2018. [Online]. Available: \url{https://github.com/scipr-lab/libsnark}
\BIBentrySTDinterwordspacing

\bibitem{app-loopring}
\BIBentryALTinterwordspacing
L.~T. Limited, ``Loopring defi: Revolutionizing decentralized finance with cutting-edge earning and trading solutions,'' 2017. [Online]. Available: \url{https://loopring.io/}
\BIBentrySTDinterwordspacing

\bibitem{asic-priormsm}
C.~Liu, H.~Zhou, P.~Dai, L.~Shang, and F.~Yang, ``Priormsm: An efficient acceleration architecture for multi-scalar multiplication,'' \emph{ACM Transactions on Design Automation of Electronic Systems}, vol.~29, no.~5, pp. 1--26, 2024.

\bibitem{gpu-batchzk}
\BIBentryALTinterwordspacing
T.~Lu, Y.~Chen, Z.~Wang, X.~Wang, W.~Chen, and J.~Zhang, ``Batchzk: A fully pipelined gpu-accelerated system for batch generation of zero-knowledge proofs,'' in \emph{Proceedings of the 30th ACM International Conference on Architectural Support for Programming Languages and Operating Systems, Volume 1}, ser. ASPLOS '25.\hskip 1em plus 0.5em minus 0.4em\relax New York, NY, USA: Association for Computing Machinery, 2025, p. 100–115. [Online]. Available: \url{https://doi.org/10.1145/3669940.3707270}
\BIBentrySTDinterwordspacing

\bibitem{gpu-cuzk}
\BIBentryALTinterwordspacing
T.~Lu, C.~Wei, R.~Yu, C.~Chen, W.~Fang, L.~Wang, Z.~Wang, and W.~Chen, ``{cuZK}: Accelerating zero-knowledge proof with a faster parallel multi-scalar multiplication algorithm on {GPUs},'' Cryptology {ePrint} Archive, Paper 2022/1321, 2022. [Online]. Available: \url{https://eprint.iacr.org/2022/1321}
\BIBentrySTDinterwordspacing

\bibitem{gpu-gzkp}
W.~Ma, Q.~Xiong, X.~Shi, X.~Ma, H.~Jin, H.~Kuang, M.~Gao, Y.~Zhang, H.~Shen, and W.~Hu, ``Gzkp: A gpu accelerated zero-knowledge proof system,'' in \emph{Proceedings of the 28th ACM International Conference on Architectural Support for Programming Languages and Operating Systems, Volume 2}, 2023, pp. 340--353.

\bibitem{zkp-sonic}
\BIBentryALTinterwordspacing
M.~Maller, S.~Bowe, M.~Kohlweiss, and S.~Meiklejohn, ``Sonic: Zero-knowledge snarks from linear-size universal and updatable structured reference strings,'' in \emph{Proceedings of the 2019 ACM SIGSAC Conference on Computer and Communications Security}, ser. CCS '19.\hskip 1em plus 0.5em minus 0.4em\relax New York, NY, USA: Association for Computing Machinery, 2019, p. 2111–2128. [Online]. Available: \url{https://doi.org/10.1145/3319535.3339817}
\BIBentrySTDinterwordspacing

\bibitem{app-ing}
\BIBentryALTinterwordspacing
E.~Morais, T.~Koens, C.~van Wijk, and A.~Koren, ``A survey on zero knowledge range proofs and applications,'' 2019. [Online]. Available: \url{https://arxiv.org/abs/1907.06381}
\BIBentrySTDinterwordspacing

\bibitem{zkp-recursive-circom}
\BIBentryALTinterwordspacing
Nalin, J.~Wang, Y.~Sun, and V.~Huang, ``Recursive zksnarks: Exploring new territory,'' 2022. [Online]. Available: \url{https://0xparc.org/blog/groth16-recursion}
\BIBentrySTDinterwordspacing

\bibitem{app-hermez}
\BIBentryALTinterwordspacing
H.~Network, ``Hermez network: Polygon hermez is an open-source zk-rollup optimised for secure, low-cost and usable token transfers on the wings of ethereum.'' 2021. [Online]. Available: \url{https://hermez.io/}
\BIBentrySTDinterwordspacing

\bibitem{nvidia-programming-guide}
\BIBentryALTinterwordspacing
NVIDIA, ``Cuda c++ programming guide,'' Online. [Online]. Available: \url{https://docs.nvidia.com/cuda/cuda-c-programming-guide/index.html}
\BIBentrySTDinterwordspacing

\bibitem{ncu-profiling-guide}
\BIBentryALTinterwordspacing
------, ``Profiling guide - nsight compute,'' Online. [Online]. Available: \url{https://docs.nvidia.com/nsight-compute/ProfilingGuide/index.html}
\BIBentrySTDinterwordspacing

\bibitem{nvidia-ampere}
\BIBentryALTinterwordspacing
------, ``Nvidia a100 tensor core gpu architecture,'' Online, 2020. [Online]. Available: \url{https://images.nvidia.com/aem-dam/en-zz/Solutions/data-center/nvidia-ampere-architecture-whitepaper.pdf}
\BIBentrySTDinterwordspacing

\bibitem{opt-pippenger}
N.~Pippenger, ``On the evaluation of powers and related problems,'' in \emph{17th Annual Symposium on Foundations of Computer Science (sfcs 1976)}, 1976, pp. 258--263.

\bibitem{opt-zprize}
\BIBentryALTinterwordspacing
A.~Pruden, ``Zprize: Accelerating the future of zero knowledge cryptography,'' 2022. [Online]. Available: \url{https://www.zprize.io/blog/announcing-zprize-results}
\BIBentrySTDinterwordspacing

\bibitem{fpga-msmac}
P.~Qiu, G.~Wu, T.~Chu, C.~Wei, R.~Luo, Y.~Yan, W.~Wang, and H.~Zhang, ``Msmac: Accelerating multi-scalar multiplication for zero-knowledge proof,'' \emph{Cryptology ePrint Archive}, 2024.

\bibitem{fpga-hardcaml}
A.~Ray, B.~Devlin, F.~Y. Quah, and R.~Yesantharao, ``Hardcaml msm: A high-performance split cpu-fpga multi-scalar multiplication engine,'' in \emph{Proceedings of the 2024 ACM/SIGDA International Symposium on Field Programmable Gate Arrays}, 2024, pp. 33--39.

\bibitem{ntt-f1}
\BIBentryALTinterwordspacing
N.~Samardzic, A.~Feldmann, A.~Krastev, S.~Devadas, R.~Dreslinski, C.~Peikert, and D.~Sanchez, ``F1: A fast and programmable accelerator for fully homomorphic encryption,'' in \emph{MICRO-54: 54th Annual IEEE/ACM International Symposium on Microarchitecture}, ser. MICRO '21.\hskip 1em plus 0.5em minus 0.4em\relax New York, NY, USA: Association for Computing Machinery, 2021, p. 238–252. [Online]. Available: \url{https://doi.org/10.1145/3466752.3480070}
\BIBentrySTDinterwordspacing

\bibitem{ntt-craterlake}
\BIBentryALTinterwordspacing
N.~Samardzic, A.~Feldmann, A.~Krastev, N.~Manohar, N.~Genise, S.~Devadas, K.~Eldefrawy, C.~Peikert, and D.~Sanchez, ``Craterlake: a hardware accelerator for efficient unbounded computation on encrypted data,'' in \emph{Proceedings of the 49th Annual International Symposium on Computer Architecture}, ser. ISCA '22.\hskip 1em plus 0.5em minus 0.4em\relax New York, NY, USA: Association for Computing Machinery, 2022, p. 173–187. [Online]. Available: \url{https://doi.org/10.1145/3470496.3527393}
\BIBentrySTDinterwordspacing

\bibitem{asic-nocap}
N.~Samardzic, S.~Langowski, S.~Devadas, and D.~Sanchez, ``Accelerating zero-knowledge proofs through hardware-algorithm co-design,'' in \emph{2024 57th IEEE/ACM International Symposium on Microarchitecture (MICRO)}, 2024, pp. 366--379.

\bibitem{zkp-cpu-perf}
S.~Samudrala, J.~Wu, C.~Chen, H.~Shan, J.~Ku, Y.~Chen, and J.~Rajendran, ``Performance analysis of zero-knowledge proofs,'' in \emph{2024 IEEE International Symposium on Workload Characterization (IISWC)}, 2024, pp. 144--155.

\bibitem{gpu-yrrid}
\BIBentryALTinterwordspacing
Y.~Software, ``Z-prize msm on the gpu submission,'' Online, October 2022. [Online]. Available: \url{https://github.com/yrrid/submission-msm-gpu}
\BIBentrySTDinterwordspacing

\bibitem{gpu-ymc}
\BIBentryALTinterwordspacing
Y.~Software and M.~Labs, ``Z-prize msm on the gpu: Yrrid / matter labs combined msm for the gpu,'' Online. [Online]. Available: \url{https://github.com/yrrid/combined-msm-gpu}
\BIBentrySTDinterwordspacing

\bibitem{ntt-rpu}
\BIBentryALTinterwordspacing
D.~Soni, N.~Neda, N.~Zhang, B.~Reynwar, H.~Gamil, B.~Heyman, M.~N.~T. Moopan, A.~A. Badawi, Y.~Polyakov, K.~Canida, M.~Pedram, M.~Maniatakos, D.~B. Cousins, F.~Franchetti, M.~French, A.~Schmidt, and B.~Reagen, ``{RPU}: The ring processing unit,'' Cryptology {ePrint} Archive, Paper 2023/465, 2023. [Online]. Available: \url{https://eprint.iacr.org/2023/465}
\BIBentrySTDinterwordspacing

\bibitem{app-zkdl}
\BIBentryALTinterwordspacing
H.~Sun, T.~Bai, J.~Li, and H.~Zhang, ``{zkDL}: Efficient zero-knowledge proofs of deep learning training,'' Cryptology {ePrint} Archive, Paper 2023/1174, 2023. [Online]. Available: \url{https://eprint.iacr.org/2023/1174}
\BIBentrySTDinterwordspacing

\bibitem{gpu-sppark}
\BIBentryALTinterwordspacing
Supranational, ``sppark: Zero-knowledge template library,'' Online, December 2022. [Online]. Available: \url{https://github.com/supranational/sppark}
\BIBentrySTDinterwordspacing

\bibitem{volkov-dissertation}
\BIBentryALTinterwordspacing
V.~Volkov, ``Understanding latency hiding on gpus,'' Ph.D. dissertation, University of California, Berkeley, {USA}, 2016. [Online]. Available: \url{https://www.escholarship.org/uc/item/1wb7f3h4}
\BIBentrySTDinterwordspacing

\bibitem{ntt-wlhm}
\BIBentryALTinterwordspacing
Z.~Wang, P.~Li, R.~Hou, and D.~Meng, ``{ NTTFusion: Efficient Number Theoretic Transform Acceleration on GPUs },'' in \emph{2023 IEEE 41st International Conference on Computer Design (ICCD)}.\hskip 1em plus 0.5em minus 0.4em\relax Los Alamitos, CA, USA: IEEE Computer Society, Nov. 2023, pp. 357--365. [Online]. Available: \url{https://doi.ieeecomputersociety.org/10.1109/ICCD58817.2023.00061}
\BIBentrySTDinterwordspacing

\bibitem{cpu-dizk}
\BIBentryALTinterwordspacing
H.~Wu, W.~Zheng, A.~Chiesa, R.~A. Popa, and I.~Stoica, ``{DIZK}: A distributed zero knowledge proof system,'' in \emph{27th USENIX Security Symposium (USENIX Security 18)}.\hskip 1em plus 0.5em minus 0.4em\relax Baltimore, MD: USENIX Association, Aug. 2018, pp. 675--692. [Online]. Available: \url{https://www.usenix.org/conference/usenixsecurity18/presentation/wu}
\BIBentrySTDinterwordspacing

\bibitem{fpga-pipemsm}
\BIBentryALTinterwordspacing
C.~F. Xavier, ``{PipeMSM}: Hardware acceleration for multi-scalar multiplication,'' Cryptology {ePrint} Archive, Paper 2022/999, 2022. [Online]. Available: \url{https://eprint.iacr.org/2022/999}
\BIBentrySTDinterwordspacing

\bibitem{zkp-orion}
\BIBentryALTinterwordspacing
T.~Xie, Y.~Zhang, and D.~Song, ``Orion: Zero knowledge proof with linear prover time,'' Cryptology {ePrint} Archive, Paper 2022/1010, 2022. [Online]. Available: \url{https://eprint.iacr.org/2022/1010}
\BIBentrySTDinterwordspacing

\bibitem{gpu-gecc}
\BIBentryALTinterwordspacing
Q.~Xiong, W.~Ma, X.~Shi, Y.~Zhou, H.~Jin, K.~Huang, H.~Wang, and Z.~Wang, ``gecc: A gpu-based high-throughput framework for elliptic curve cryptography,'' 2024. [Online]. Available: \url{https://arxiv.org/abs/2501.03245}
\BIBentrySTDinterwordspacing

\bibitem{roofline-hierarchy}
\BIBentryALTinterwordspacing
C.~Yang, T.~Kurth, and S.~Williams, ``Hierarchical roofline analysis for gpus: Accelerating performance optimization for the nersc-9 perlmutter system,'' \emph{Concurr. Comput. Pract. Exp.}, vol.~32, no.~20, 2020. [Online]. Available: \url{https://doi.org/10.1002/cpe.5547}
\BIBentrySTDinterwordspacing

\bibitem{fpga-falic}
Y.~Yang, Z.~Lu, J.~Zeng, X.~Liu, X.~Qian, and Z.~Yu, ``Falic: An fpga-based multi-scalar multiplication accelerator for zero-knowledge proof,'' \emph{IEEE Transactions on Computers}, 2024.

\bibitem{zeus-energy}
J.~You, J.-W. Chung, and M.~Chowdhury, ``Zeus: Understanding and optimizing {GPU} energy consumption of {DNN} training,'' in \emph{USENIX NSDI}, 2023.

\bibitem{app-zcash}
\BIBentryALTinterwordspacing
ZCash, ``Zcash - internet money.'' [Online]. Available: \url{https://z.cash/}
\BIBentrySTDinterwordspacing

\bibitem{app-zombie}
\BIBentryALTinterwordspacing
C.~Zhang, Z.~DeStefano, A.~Arun, J.~Bonneau, P.~Grubbs, and M.~Walfish, ``Zombie: Middleboxes that {Don{\textquoteright}t} snoop,'' in \emph{21st USENIX Symposium on Networked Systems Design and Implementation (NSDI 24)}.\hskip 1em plus 0.5em minus 0.4em\relax Santa Clara, CA: USENIX Association, Apr. 2024, pp. 1917--1936. [Online]. Available: \url{https://www.usenix.org/conference/nsdi24/presentation/zhang-collin}
\BIBentrySTDinterwordspacing

\bibitem{zkp-virgo}
\BIBentryALTinterwordspacing
J.~Zhang, T.~Xie, Y.~Zhang, and D.~Song, ``Transparent polynomial delegation and its applications to zero knowledge proof,'' Cryptology {ePrint} Archive, Paper 2019/1482, 2019. [Online]. Available: \url{https://eprint.iacr.org/2019/1482}
\BIBentrySTDinterwordspacing

\bibitem{asic-pipezk}
Y.~Zhang, S.~Wang, X.~Zhang, J.~Dong, X.~Mao, F.~Long, C.~Wang, D.~Zhou, M.~Gao, and G.~Sun, ``Pipezk: Accelerating zero-knowledge proof with a pipelined architecture,'' in \emph{2021 ACM/IEEE 48th Annual International Symposium on Computer Architecture (ISCA)}, 2021, pp. 416--428.

\bibitem{fpga-bstmsm}
B.~Zhao, W.~Huang, T.~Li, and Y.~Huang, ``Bstmsm: A high-performance fpga-based multi-scalar multiplication hardware accelerator,'' in \emph{2023 International Conference on Field Programmable Technology (ICFPT)}, 2023, pp. 35--43.

\bibitem{gpu-elasticmsm}
\BIBentryALTinterwordspacing
X.~Zhu, H.~He, Z.~Yang, Y.~Deng, L.~Zhao, and R.~Hou, ``Elastic {MSM}: A fast, elastic and modular preprocessing technique for multi-scalar multiplication algorithm on {GPUs},'' Cryptology {ePrint} Archive, Paper 2024/057, 2024. [Online]. Available: \url{https://eprint.iacr.org/2024/057}
\BIBentrySTDinterwordspacing

\bibitem{ntt-ozcan}
\BIBentryALTinterwordspacing
A.~Şah Özcan and E.~Savaş, ``Two algorithms for fast {GPU} implementation of {NTT},'' Cryptology {ePrint} Archive, Paper 2023/1410, 2023. [Online]. Available: \url{https://eprint.iacr.org/2023/1410}
\BIBentrySTDinterwordspacing

\end{thebibliography}

\end{document}